\documentclass[twocolumn,floatfix,aps,prd,showpacs,footinbib,amsmath,amssymb,amsfonts]{revtex4}
\usepackage{bm}

\usepackage{graphicx}

\usepackage[dvips,usenames]{color}
\usepackage{dsfont}
\usepackage{hyperref}

\newcommand{\be}{\begin{equation}}
\newcommand{\ee}{\end{equation}}
\newcommand{\bea}{\begin{eqnarray}}
\newcommand{\eea}{\end{eqnarray}}
\newcommand{\gr}[1]{\bm{#1}}

\newcommand{\ii}{\mathrm{i}}

\def\st{\sigma_{\mathrm T}}

\def\dd{{\rm d}}
\def\HH{\mathcal{H}}

\definecolor{Myblue}{rgb}{0,0,0}
\definecolor{Myred}{rgb}{0,0,0}
\def\iB{{\color{Myred}i}}
\def\jB{{\color{Myred}j}}
\def\kB{{\color{Myred}k}}
\def\lB{{\color{Myred}\ell}}
\def\bB{{\color{Myred}\nu}}
\def\zT{{\underline{0}}}
\def\iT{{\underline{i}}}
\def\jT{{\underline{j}}}
\def\kT{{\underline{k}}}
\def\lT{{\underline{\ell}}}

\def\aT{{\underline{a}}}
\def\bT{{\underline{b}}}
\def\cT{{\underline{c}}}
\def\dT{{\underline{d}}}

\def\ipl{{\rm f}}
\def\ib{{\mathrm b}}
\def\ir{{\mathrm r}}
\def\ie{{\mathrm e}}
\def\ip{{\mathrm p}}
\def\is{{\mathrm s}}

\def\etaC{{\eta_{\rm C}}}

\newcommand{\eq}{{\rm eq}}

\begin{document}
\title{The seed magnetic field generated during recombination}

\author{Elisa Fenu$^1$}
\author{Cyril Pitrou$^2$}
\author{Roy Maartens$^{2,3}$}
\affiliation{ $^1\,$Universit\'e
de Gen\`eve, D\'epartement de Physique Th\'eorique,
CH-1211 Gen\`eve, Switzerland\\
$^2\,$Institute of Cosmology \& Gravitation,
University of Portsmouth, Portsmouth PO1 3FX, United Kingdom\\
$^3\,$Department of Physics, University of Western Cape, Cape Town
7535, South Africa}


\date{\today}

\begin{abstract}

Nonlinear dynamics creates vortical currents when the
tight-coupling approximation between photons and baryons breaks
down around the time of recombination. This generates a magnetic
field at second order in cosmological perturbations, whose power
spectrum is fixed by standard physics, without the need for any ad
hoc assumptions. We present the fully general relativistic calculation of
the magnetic power spectrum, including the effects of metric
perturbations, second-order velocity and photon anisotropic stress, thus
generalizing and correcting previous results. We also show that
significant magnetogenesis continues to occur after recombination.
The power spectrum $\sqrt{k^3 P_B} $ 
decays as $k^4$ on large scales, and 
grows as $k^{0.5}$ on small scales,
down to the limit of our numerical computations, $\sim 1\,$Mpc. On cluster scales, the created field has strength $\sim 3\times 10^{-29}\,$Gauss.

\end{abstract}

\maketitle

\section{Introduction}

Evidence is growing for magnetic fields on larger and larger
scales in the Universe (see e.g. the reviews
\cite{Giovannini:2003yn,Subramanian:2008tt}). In galaxies, the
fields have strength of order $\mu$Gauss, ordered on scales $\sim
1-10\,$kpc. Fields of strength $\sim 1-10^{-2}\mu$G on scales
$\sim 0.1-1\,$Mpc have been detected in galaxy clusters, and there
is evidence of magnetic fields in superclusters. Recently, new
evidence has been presented for intergalactic magnetic fields:
high energy gamma-rays from distant sources can initiate
electromagnetic pair cascades when interacting with the
extragalactic photon background; the charged component of the
cascades will be deflected by magnetic fields, affecting the
images of the sources. Using observations from FERMI, a lower
bound
of order $10^{-16}\,$G
has been claimed for the strength of
fields in the filaments and voids of the cosmic web
\cite{2010ApJ...722L..39A,Neronov:1900zz,2010arXiv1009.1782D,Essey:2010nd}.

The origin of these fields is still unclear 
(see e.g. \cite{Brandenburg:2004jv,Kulsrud:2007an,Kandus:2010nw}). 
They could have been
generated via astrophysical processes during 
the nonlinear collapse stage of 
structure
formation. There remain unresolved difficulties in explaining how
these astrophysical seed fields lead to fields of the observed
strength and coherence scales. Alternatively, the fields could be
primordial seed fields -- created in the very early Universe,
during inflation, or during subsequent phase transitions. In
principle inflation can generate fields on all scales -- but
unknown physics must be invoked to achieve non-minimal coupling of
the electromagnetic field. The electroweak and QCD transitions
can only produce fields on very small scales, up to the Hubble
radius at magnetogenesis (and their amplitude is strongly
constrained by their gravitational wave production before
nucleosynthesis \cite{Caprini:2009pr}).

Primordial magnetogenesis also takes place in the cosmic plasma
after particle/anti-particle annihilation. This avoids the problem
of exotic physics that faces inflationary magnetogenesis --
standard Maxwell theory and standard cosmological perturbations in
the cosmic plasma inevitably lead to magnetic fields. It also
avoids the small coherence scale problem facing electroweak and
QCD fields. However, the problem is the weakness of the fields,
since this effect occurs at second and higher order in
cosmological perturbations.

The key question is how weak is the field and how does it vary with scale? Differing 
qualitative estimates of the field strength have been given by
\cite{Hogan:2000gv,Berezhiani:2003ik,Gopal:2004ut,Siegel:2006px,Kobayashi:2007wd,Maeda:2008dv}.
The power spectrum was first 
numerically computed by \cite{Matarrese2004}, which
differs significantly from
ours. More recently, \cite{Ichiki:2007hu} presented a power
spectrum that is closer to our result. We discuss below the differences between previous results and ours. Our analysis is
the first complete general relativistic computation of the power spectrum, taking into account all effects.

\begin{figure*}[!htp]
\includegraphics[width=8.5cm]{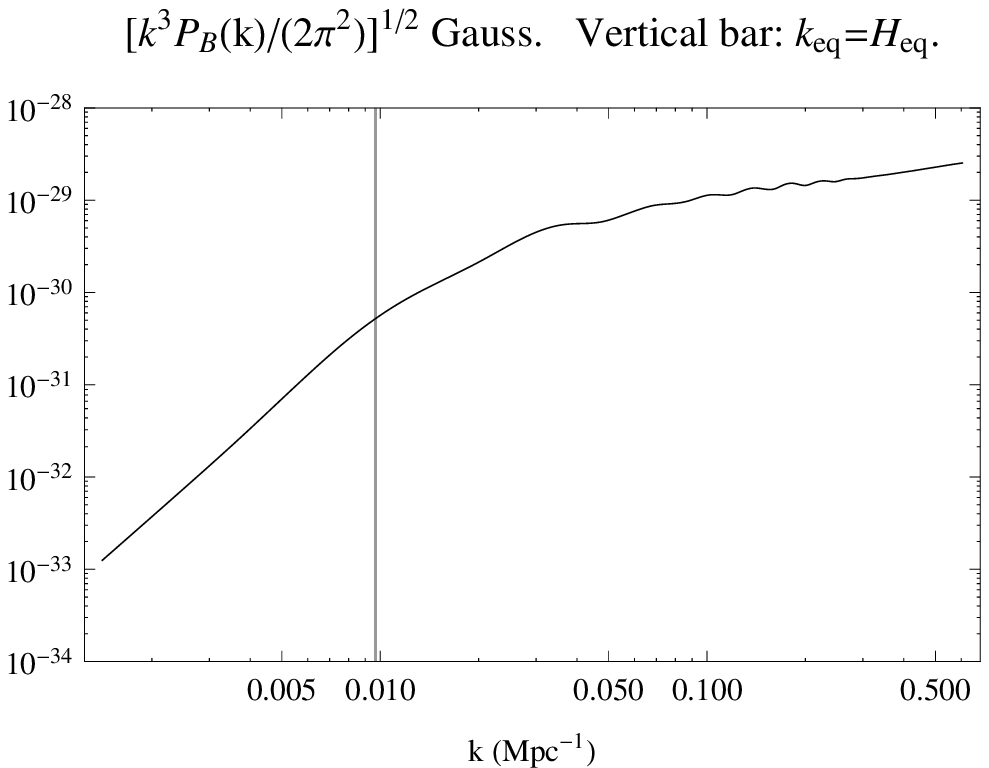}\quad
\includegraphics[width=8.5cm]{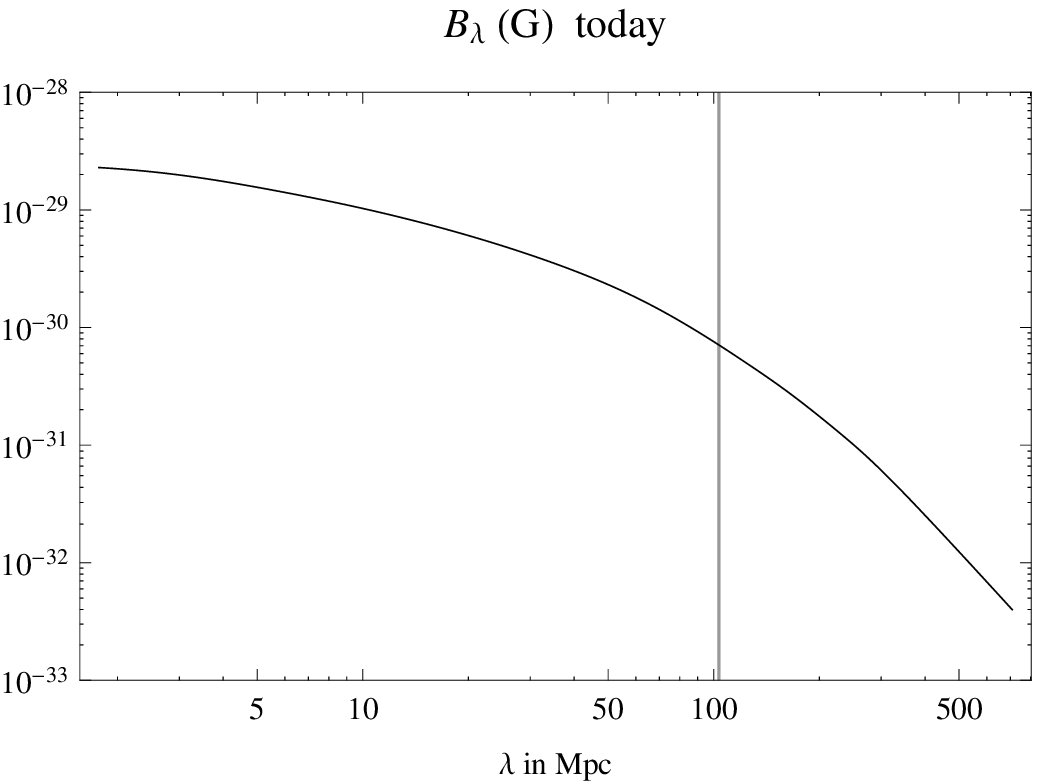}
\caption{{\em Left:} Magnetic field spectrum today.
{\em Right:} Comoving magnetic field strength today at a given scale.} \label{fig1one}
\end{figure*}

Our result is shown in 
Fig. \ref{fig1one}. The power spectrum behaves as
 \be
\sqrt{k^3 P_B} \propto \left\{ \begin{array}{ll} k^4 & k\ll k_{\rm eq} \\ k^{0.5} & k \gg k_{\rm eq}\,. \end{array} \right.
 \ee
On cluster scales the
comoving field strength is
 \be\label{b1mpc}
B_{1\,{\rm Mpc}} \sim 3\times 10^{-29}\,{\rm G}.
 \ee
 
Thus the field generated around recombination is too weak to act
as a seed for the observed field strength of order $\mu$G. Adiabatic contraction of the
magnetic flux lines during 
nonlinear collapse of structures provides an
enhancement of $\sim 10^3$, while the nonlinear dynamo mechanism
has an amplification factor $\sim 10^{8}$ (with many remaining
uncertainties). 
Note that hydrodynamical and turbulence effects during nonlinear collapse themselves generate a field of order $10^{-20}\,$G -- which is also too small to account for the observed galactic and cluster fields \cite{Kulsrud:2007an}.

The field (\ref{b1mpc}) is also too weak to imprint detectable
effects on the CMB. Nevertheless it is a real property of the
standard cosmological model, and may have some impact on early
structure formation during the `dark ages' if it is the only
primordial field. 
(See e.g. \cite{Sethi:2009dd,Schleicher:2010ph} for the role of magnetic fields in structure formation during the dark ages.)

As shown below, the magnetic field is given by
 \bea
\label{Maxwell-0} \left(a^2 B^i\right)'&=&-a^2
\epsilon^{ijk}\partial_{j}\Big[\left(1+\Phi-\Psi \right)E_k
\Big],\\
\label{GenerationE-0}
E^i &\approx&  -\frac{4\rho_\gamma \st}{3 e}\Big(\Delta
v_{\ib\gamma}^i +\frac{2}{5}\Theta^i_jv_{\ib}^j\Big),
 \eea
where $\Phi,\Psi$ are first-order metric perturbations, $\Delta
v_{\ib\gamma}^i=v_{\ib}^i - v_{\gamma}^i$ is the photon-baryon
velocity difference, and $ \Theta^i_j$ is the photon quadrupole moment, from anisotropic stress. This leads to three types of source terms for magnetogenesis:
 \bea \label{b-sources}
\left(a^2 B\right)'&=& S_1\big[ \Delta v_{\ib\gamma}^{(2)}\big]+
S_2\big[\big\{\delta_\gamma^{(1)}+\Phi^{(1)}-\Psi^{(1)} \big\}\Delta v_{\ib\gamma}^{(1)} \big] \nonumber\\ &&~~ + S_3\big[\Theta_\gamma^{(1)} v_{\ib}^{(1)}\big].
 \eea
The first source term is second-order, while the other two are
quadratic in first-order quantities.
The contributions of the source terms to the power spectrum are shown in Fig. \ref{fig1} (left).

Our paper builds on the physical analysis of nonlinear plasma
dynamics presented in
\cite{Maartens:1998xg,Matarrese2004,Ichiki:2007hu,Kobayashi:2007wd,Takahashi:2007ds,Maeda:2008dv,Pitrou2008}.
The key features of the dynamics are as follows.
\begin{itemize}\itemsep=-4pt
\item
The electric field ensures that the proton-electron relative
velocity is always strongly suppressed in comparison with the
photon-electron relative velocity -- even at high energies when the Compton
interaction is stronger than the Coulomb interaction.
\item
Vorticity induced in the electron fluid is thus transferred
almost
entirely to the protons, and the baryon vorticity evolution is
determined by the two-fluid dynamics of photons and baryons, which
is very close to the equations of CMB dynamics. We use the
second-order Boltzmann code of \cite{Pitrou2008}.
 \item
The limit $v_{\rm e} - v_{\rm \gamma}\to 0$ and  $v_{\rm
p}-v_{\rm e}\to 0$  is not equivalent to setting $v_{\rm p}=v_{\rm
e}=v_{\rm \gamma}$ in the momentum exchange equations, and the
limit must be taken consistently.
 \item
At first order, cosmological vector perturbations are zero after
inflation, in the standard model. Magnetogenesis requires vortical
currents, and these can therefore only be generated at second
order, via mode-mode coupling of first-order scalar perturbations.
This remains true even in the presence of topological defects,
which are active sources for vector perturbations: at first order,
the vector perturbations induced by the defects cannot break
vorticity conservation in the cosmic plasma
\cite{Hollenstein:2007kg}.
 \item
On large scales there is some cancellation amongst the source
terms in  (\ref{b-sources}) (this is evident from Fig. \ref{fig1}). Neglecting any of the effects
can thus lead to unreliable results.
\item
The magnetic field continues to be created after recombination,
due to the residual nonzero ionization fraction. If the numerical
integration is stopped at recombination, then the comoving field
is under-estimated by a factor $\sim 10$ (see Fig. \ref{fig1}).
\end{itemize}

The plan of the paper is as follows. In the next section we review
and clarify the magnetic and electric field generation beyond the
tight-coupling limit. In Sec.~\ref{results}, we detail the
numerical integration of the differential evolution equations at
second order in cosmological perturbations that we perform in
order to solve for the magnetic field spectrum. We also provide
analytical insight into the time and scale behaviors of the
numerical results. We compare our results with previous work in
Sec.~\ref{SecComparison}. Details of some calculations are given
in the Appendices.

\section{Understanding the origin of the magnetic field}
\label{SecMagneticField}

\subsection{Interactions in the cosmic plasma}

The stress-energy tensor of a species $\is$ satisfies
 \be
\nabla_{\nu}T_{\is}^{\mu \nu}=\sum_{\rm r} C^\mu_{\is {\rm r}}\,,
~~\sum_\is \nabla_{\nu}T_{\is}^{\mu \nu}=0\,, \label{stress
energy}
 \ee
where $C^\nu_{\is {\rm r}}(= -C^\nu_{{\rm r} \is})$ encodes all
the effects of interactions with species $r$. Relative to
observers with $4$-velocity $u^\mu$, the energy density transfer
rate is $-u_\mu C^\mu_{\is {\rm r}}$ and the momentum density
transfer rate is $C^{\mu\perp}_{\is {\rm r}}=
h^\mu_{\nu}C^\nu_{\is {\rm r}} $, where the projector
is $h_{\mu}^{\nu}\equiv
\delta_\mu^{\nu}+u_\mu u^\nu$.

The Euler equation for a species $\is$ is given in general by
 \be \label{geneul}
\nabla_\nu T_\is^{\nu\mu\perp}= \sum_\ir C^{\mu\perp}_{\is\ir}\,.
 \ee

The kinematics of
$u^\mu$ are described by decomposing its covariant derivative as \cite{Maartens:1998xg,Tsagas2007}
 \be
\nabla_\mu u_\nu = {1\over3}\theta h_{\mu \nu}+\sigma_{\mu \nu}+
\omega_{\mu\nu}- u_\mu \dot{u}_\nu \,,
 \ee
where $\theta$ is the volume expansion, $\sigma_{\mu \nu}$ is the
projected (i.e. orthogonal to $u^\mu$), symmetric and tracefree
shear, $\omega_{\mu\nu}$ is the projected antisymmetric vorticity,
and $\dot u_\mu=u^\nu \nabla_\nu u_\mu $ is the projected
acceleration. The vorticity vector is defined as
 \be \label{vortdef}
\omega_\nu\equiv \epsilon_{\mu \nu \lambda} \omega^{\nu
\lambda}\,,~~ \epsilon_{\mu \nu \lambda}\equiv u^\tau
\epsilon_{\tau\mu \nu \lambda}\,,
 \ee
where the totally antisymmetric tensor is defined by
$\epsilon_{0123}=\sqrt{-g}$. (Note that our sign convention for $\omega_{\mu\nu}$ and definition of $\omega_\mu$ recover the  Newtonian limit, and differ from \cite{Maartens:1998xg,Tsagas2007}.)

In the period of interest, from the end of particle/anti-particle
annihilation up to now ($T_\gamma \lesssim 500 {\rm keV}$, $z
\lesssim 2\times 10^9$), the relevant species are protons,
electrons, photons, and when recombination occurs, hydrogen atoms.
Neutrinos affect only the
background dynamics and the gravitational potentials in the
Einstein equations. The Faraday tensor of the electromagnetic
field defines electric and magnetic fields  measured by $u^\mu$
observers:
 \be \label{ebfara}
E^\mu = F^{\mu \nu}u_{\nu},~ B^\mu = \frac{1}{2}\epsilon^{\mu \nu
\lambda} F_{\nu \lambda} \,.
 \ee
Protons and electrons couple to the electromagnetic field through
the term $C^\mu_{\is F}= F^\mu_{\phantom{\mu} \nu} j_\is^\nu$, where
$s=\ip,\ie$ and $j_s^\nu$ is the electric 4-current. Then
$\nabla_\nu T^{\mu\nu}_F= -\sum_\is F^\mu_{\phantom{\mu}
\nu}j_\is^\nu$. We have $j_\is^\mu=q_\is n_\is u_\is^\mu$, where $q_\is$ is the
particle charge, $n_\is$ is the number density (in the rest frame)
and the 4-velocity of species $\is$ is
 \be \label{svel}
u^{\mu}_\is=\gamma_\is (u^\mu +v_\is^\mu), ~~ u_\mu v_\is^\mu=0,~~
\gamma_\is = \left(1-v_\is^2 \right)^{-1/2}.
 \ee
Here $\gamma_\is v_\is^\mu$ is the relative velocity of $\is$ measured
by $u^\mu$. Maxwell's equations are given in Appendix \ref{maxeq}.

The momentum transfer rates are given by
 \bea
\hspace*{-2em} \label{Collterms1}
&&C^{\mu\perp}_{\ip \ie}= -e^2 n_{\ie} n_{\ip} 
\etaC \Delta v^\mu_{\ip \ie}\,, ~~ \Delta v^\mu_{\ip\ie} \equiv
\gamma_{\ip}v_{\ip}^\mu- \gamma_{\ie}v_{\ie}^\mu \,,\\
\hspace*{-2em} \label{Collterms2}&& C^{\mu\perp}_{\ie \gamma}=
-\frac{4}{3}
n_{\ie} \rho_{\gamma}\st \Big(\Delta v_{\ie\gamma}^\mu
+\frac{2}{5}\Theta^{\mu}_{\nu}v_{\ie}^\nu \Big),\\
\hspace*{-2em} \label{Collterms3}&&C^{\mu\perp}_{\ip \gamma}=
-\frac{4}{3}
\beta^2 n_{\ip} \rho_{\gamma}\st \Big(\Delta v_{\ip\gamma}^\mu
+\frac{2}{5}\Theta^{\mu}_{\nu}v_{\ip}^\nu \Big),~
\beta \equiv {m_{\ie} \over m_{\ip}},\\
\hspace*{-2em} \label{Collterms4}&&C^{\mu\perp}_{\is F}=
q_\is n_{\is}
\big( E^\mu+\epsilon_{\mu\nu\tau}v_\is^\nu B^\tau
\big),~\is=\ie,\ip\,.
 \eea
The radiation energy density  $\rho_\gamma$, the quadrupole of the
radiation temperature anisotropy $\Theta_{\mu\nu}$, and the number
densities $n_\is$, are as measured by $u^\mu$ observers. In the rest
frame $u_\is^\mu$, the electrons and protons are well approximated
by pressure-free matter, $T_\is^{\mu\nu}=\rho_\is^{\rm rest}u_\is^\mu
u_\is^\nu$, where $\rho_\is^{\rm rest}$ is the rest-frame density
measured by $u_\is^\mu$. In the $u^\mu$ frame, there is effective
pressure, momentum density and anisotropic stress:
$T_\is^{\mu\nu}=\rho_\is u^\mu u^\nu +P_\is h^{\mu\nu} +2q_\is^{(\mu}
u^{\nu)}+\pi_\is^{\mu\nu}$, where \cite{Maartens:1998xg},
 \bea\label{dyntrans1}
&& \rho_\is\equiv m_\is n_\is=\gamma_\is^2\rho_\is^{\rm rest}\,, ~~ P_\is=
{1\over3} v_\is^2\rho_\is\,, \\ \label{dyntrans2} && q_\is^\mu=\rho_\is
v_\is^\mu\,, ~~ \pi_\is^{\mu\nu}=\rho_\is \Big( v_\is^\mu v_\is^\nu -
{1\over3} v_\is^2 h^{\mu\nu} \Big).
 \eea
The Thomson cross-section is $\st=8\pi\alpha^2/(3m_\ie^2)$, and
the Coulomb interaction is governed by the electrical resistivity
 \bea
\etaC &=& \frac{\pi e^2 \sqrt{m_\ie} \ln \Lambda}{T^{3/2}}
\nonumber\\
&\simeq & 10^{-12}{\rm sec}\left(
\frac{1+z}{10^3}\right)^{-3/2} \left(\frac{\ln
\Lambda}{10}\right),
 \eea
where $\Lambda$ is the Coulomb logarithm. On cosmological time
scales the magnetic field diffuses below a length scale $\sim
\sqrt{\etaC/H_0}\sim 100\,$AU, so that diffusion can safely be
ignored \cite{Ichiki:2007hu}. The characteristic time scales for
electrons interacting via the Coulomb and Thomson interactions are
 \bea
\hspace*{-2em}&& \tau_{\rm C}=\frac{m_\ie}{e^2 n_\ie \etaC} \simeq
\frac{20 \,{\rm sec}}{x_\ie}\left( \frac{1+z}{10^3}\right)^{\!\!-3/2}\!,  x_\ie \equiv
{ n_\ie \over n_\ie+n_{\rm H}}, \label{tauc}
  \\ \hspace*{-2em}
&&\tau_{\rm T}=\frac{m_\ie}{\st \rho_\gamma} \simeq
5\times 10^{8} {\rm sec} \left( \frac{1+z}{10^3}\right)^{\!\!-4} \!,\label{taut}
 \eea
where $n_\ie$ is the number density of free electrons and $x_\ie$ is the fraction of free electrons.
We used $n_{\ie 0}+n_{{\rm H}0}\simeq
3\times 10^{-7}\,$cm$^{-3}$ \cite{Takahashi:2008gn}.
The time scale which characterizes the evolution of the plasma
can be taken as
 \bea \hspace*{-1em}
\tau_{\rm evo}(z)&=& {\rm min}\, \{ \tau_{\rm S}(z),\, \tau_{1\,{\rm Mpc}}(z) \} \nonumber\\\hspace*{-1em} &=& {\rm min}\, \Big\{{1 \over \sqrt{ H(z)\st n_\ie(z)}},\, {1 \over (1+z)k_{1\,{\rm Mpc}}} \Big\}.
 \eea
Here $\tau_{\rm S}$ is the Silk damping time and $1\,$Mpc is taken as the minimum comoving scale on which we can trust a second-order perturbative analysis up to redshift $z=0$.

\subsection{Electric field}

The Euler equation (\ref{geneul}) for the proton and electron velocities is
given by \cite{Maartens:1998xg}:
 \be
m_\is n_\is\Big(\dot{v}_\is^{\mu\perp}+\dot{u}^\mu + K_\is^\mu \Big)=
C^{\mu\perp}_{\is {\rm r}} + C^{\mu\perp}_{\is\gamma}+
C^{\mu\perp}_{\is F},
\label{pe}
 \ee
where $\is,{\rm r}= \ip, \ie$ and
 \bea
K_\is^\mu &=& \Big( {\dot{n}_\is \over n_\is}+{4\over3}\theta+
\dot{u}_\nu v_\is^\nu+ {1\over n_\is}v_\is^\nu {\rm D}_\nu n_\is +{\rm
D}_\nu v_\is^\nu \Big)v_\is^\mu \nonumber\\ &&~+
\left(\sigma^\mu{}_\nu- \omega^\mu{}_\nu \right) v_\is^\nu + v_\is^\nu
{\rm D}_\nu v_\is^\mu.\label{euler-cov}
 \eea
The covariant spatial derivative ${\rm D}_\mu$ is defined in
(\ref{deriv}). The first term on the right of (\ref{euler-cov}) describes not only the evolution due
to the expansion of the universe which conserves the particles, but also the evolution of the number
density due to recombination which does not conserve the particles
when hydrogen atoms are formed around recombination.

From now on we expand in perturbations around a Friedmann
background, up to second order. The metric in Poisson gauge is
 \bea
    \label{metric}
\dd s^2= a^2 \left[-(1 + 2\Phi )\dd\eta^2 + 2 S_{\iB} \dd
x^{\iB}\dd\eta +  (1-2 \Psi)\dd \bm{x}^2\right]
 \eea
where $S_{\iB}$ is a vector perturbation ($\partial^\iB S_\iB=0$)
and enters only at second order. Perturbed quantities are expanded
according to $X=\bar X + X^{(1)}+X^{(2)}$. Only the first order of
scalar perturbations $\Phi$ and $\Psi$ will enter the evolution
equation of the magnetic field, so we omit the superscripts for
them. The explicit form of the term $\dot{v}_\is^{\mu\perp}+\dot{u}^\mu + K_\is^\mu$
in (\ref{pe}) is then given by (\ref{Eulerappendix}), with $w_\is=0= c_\is^2$.

We set
$n_\ie=n_\ip\equiv n$, since we find that the final expression of
the resulting electric field is not affected by $n_\ie-n_\ip$, in
agreement with \cite{Takahashi:2007ds}.

In order to obtain a dynamical equation for the velocity
difference $\Delta v_{\ip\ie}^\mu = v_\ip^\mu -v_\ie^\mu$, we use
(\ref{pe}) to obtain
 \bea
 m_\ie  n  \Big(\Delta \dot{v}_{\ip\ie}^{\mu\perp}+\Delta
K_{\ip\ie}^\mu \Big)&=& (1+\beta)en E^\mu  +
C^{\mu\perp}_{\ip\ie}- C^{\mu\perp}_{\ie\gamma} \nonumber\\  &&~+
\beta\left(C^{\mu\perp}_{\ip \ie} +
C^{\mu\perp}_{\ip\gamma}\right)  \,. \label{delvdot}
 \eea
The Lorentz force term in (\ref{Collterms4}) has been neglected
since it is higher order. We define the baryon velocity as the
velocity of the centre of mass of the charged particles; then
 \bea
\hspace*{-2em}
&&(m_\ip +m_\ie) v^\mu_\ib = m_\ip v^\mu_\ip +m_\ie v^\mu_\ie~,\\
\hspace*{-2em} && v^\mu_\ip= v^\mu_\ib +\frac{\beta}{1+\beta}
\Delta v_{\ip\ie}^\mu~, \quad v^\mu_\ie= v^\mu_\ib
-\frac{1}{1+\beta} \Delta v_{\ip\ie}^\mu~.\label{bpevel}
 \eea
In principle, the baryon velocity can be different from the velocity
of hydrogen, i.e. of electrons and protons recombined, but thermal
collision ensure that hydrogen atoms follow closely the electrons and protons.

Using (\ref{delvdot})--(\ref{bpevel}) and the explicit forms
(\ref{Collterms1})--(\ref{Collterms3}) of the collision terms, we
obtain
 \bea \hspace*{-1.5em}
&& m_\ie  \Big(\Delta \dot{v}_{\ip\ie}^{\mu\perp}+\Delta
K_{\ip\ie}^\mu \Big) = (1+\beta)e E^\mu -(1+\beta) e^2n\etaC
\Delta v_{\ip\ie}^\mu  \nonumber \\ \hspace*{-1.5em} &&~~~~~~~~~~
+\frac{4}{3}\st \rho_\gamma
\Big[ (1-\beta^3)\Big(\Delta v_{\ib\gamma}^\mu
+\frac{2}{5}\Theta^{\mu}_{\nu}v_{\ib}^\nu\Big) \nonumber \\
\hspace*{-1.5em} &&~~~~~~~~~~~~
-\frac{1+\beta^4}{1+\beta}\Big(\Delta v_{\ip\ie}^\mu
+\frac{2}{5}\Theta^{\mu}_{\nu}\Delta
v_{\ip\ie}^\nu\Big)\Big].\label{evolEq}
 \eea
We show below that the $\Theta^{\mu}_{\nu}\Delta v_{\ip\ie}^\nu$
term can be neglected, since it is higher order.

Equation (\ref{evolEq}) shows that an electric field can be
generated by nonzero velocity differences $\Delta v_{\ip\ie}$ and
$\Delta v_{\gamma\ib}$. The Maxwell equation (\ref{MaxwellBdot})
shows that then $B^\mu$ can be generated, provided that $E^{\mu}$
is transverse. We will show that the generated electric field
keeps electrons and protons more bound together and therefore
leads to a decrease in $\Delta v_{\ip\ie}$, which becomes
negligible compared to $\Delta v_{\gamma \ie}$.

Neglecting third
order terms, the Maxwell equation (\ref{MaxwellE}) can be
rewritten in terms of the velocity difference $\Delta
v_{\ip\ie}^\mu$ as
\be\label{Eqdvep}
\Delta v_{\ip\ie}^\mu=\frac{1}{e n  } \Big({\rm curl}\, B^\mu-
\dot{E}^{\mu\perp}-{2\over3}\theta E^\mu+ \sigma^{\mu\nu}E_\nu
\Big),
 \ee
where we used (\ref{chnoden}).

In order to estimate the magnitudes of the various contributions
in the stationary regime, we expand all evolving quantities in
frequency space:
 \be
M^\mu(\bm{ x},\eta) = \int_0^\infty \dd \omega \hat{M}^\mu(\bm{
x},\omega) {\rm e}^{\ii\omega \eta}~,
 \ee
where the mode $\hat{M}^\mu$ has characteristic oscillation
frequency $\omega\simeq \tau_{\rm evo}^{-1}$. In terms of the
characteristic timescales (\ref{tauc}) and (\ref{taut}), we find from (\ref{evolEq}) and (\ref{Eqdvep}) that
 \bea \hspace*{-2em}
&&\hat{E}^\mu \Big[(1+\beta) + {\cal
O}\Big(\frac{\etaC\tau_{\rm C}} {\tau_{\rm
evo}^2}+\ii\frac{4}{3} \frac{\etaC} { \tau_{\rm evo}}
+\ii \frac{\etaC\tau_{\rm C}}
{\tau_{\rm evo} \tau_{\rm T}}\Big)\Big]\nonumber\\
&&~~~=\eta_{\rm C,eff} \Big[(1+\beta)+ {\cal
O}\Big(\ii\frac{\eta_{\rm C}\tau_{\rm C}}{\eta_{\rm C,eff} \tau_{\rm evo}}\Big)\Big]{\rm curl}\, \hat{B}^\mu \nonumber\\
&&~~~~~ -\frac{4m_\ie}{3e\tau_{\rm
T}}(1-\beta^3)\Big[{\Delta \hat{v}}_{\ib\gamma}^\mu
+\frac{2}{5}\Theta^{\mu}_{\nu}\hat{v}_{\ib}^\nu\Big],
\label{masterEquationforE}
 \eea
where we used $\Delta K_{\ip\ie} = {\cal O}(\Delta \dot v_{\ip\ie})$, and we defined \cite{Ichiki:2007hu}
 \be
\eta_{\rm C,eff}\equiv \etaC
\Big[1+\frac{4(1+\beta^4)}{3(1+\beta)^2} \frac{\tau_{\rm C}}{\tau_{\rm T}} \Big].
 \ee
\begin{figure}[!htp]
\includegraphics[width=8.5cm]{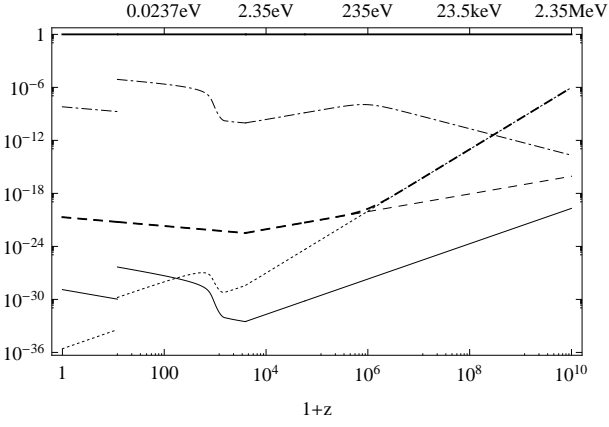}
\caption{Evolution with redshift of different ratios between
characteristic times that arise in (\ref{masterEquationforE}), compared with unity (thick black line):
$\etaC\tau_{\rm C}/(\tau_{\rm evo}^2)$ (thin solid),
$\etaC/ \tau_{\rm evo}$ (thin dashed), $\etaC\tau_{\rm
C}/(\tau_{\rm evo} \tau_{\rm T})$ (dotted), $\etaC\tau_{\rm
C}/ (\eta_{\rm C,eff}\tau_{\rm evo})$ (dot-dashed) and
$\eta_{\rm C,eff}/\tau_{\rm evo}$ (thick dashed).
(The jumps in the curves occur at reionization.)
}
\label{fig:times}
\end{figure}

Given the hierarchy of the different timescales involved in
(\ref{masterEquationforE}), it follows that the largest
contribution to the resulting electric field is given by the
velocity difference $\Delta v_{\ib\gamma}^\mu$. This can be seen
in Fig.~\ref{fig:times}, where we plot the different ratios of
typical timescales that enter in (\ref{masterEquationforE}).
Specifically, all the plotted ratios are always well below unity
for the period of interest, from very large redshift until today,
even accounting for recombination around $z\simeq 1080$. This
allows us to write
 \bea
 E^\mu  &\simeq&
\eta_{\rm C,eff}\,{\rm curl}\,  B^\mu \nonumber\\
&&{} -\frac{4 m_\ie}{3 e\tau_{\rm
T}}\frac{1-\beta^3}{1+\beta}\Big(\Delta v_{\ib\gamma}^\mu
+\frac{2}{5}\Theta^{\mu}_{\phantom{\mu}\nu}
v_{\ib}^\nu\Big).\label{mezzobis}
 \eea
In order to compute the final magnetic field produced by such an
electric field, we consider the curl of the electric field,
governed by Maxwell's equation (\ref{MaxwellBdot}). In frequency
space
 \bea\label{EqBroughly} \ii\frac{\hat{B}^\mu}{\tau_{\rm
evo}} &\simeq& -\eta_{\rm C,eff}\,
{\rm curl}\,{\rm curl}\,\hat{B}^\mu  \nonumber \\
&&{}+\frac{4 m_\ie}{3 e \tau_{\rm T}}\frac{1-\beta^3}{1+\beta} {\rm
curl}\Big({\Delta \hat{v}}_{\ib\gamma}^\mu
+\frac{2}{5}\Theta^{\mu}_{\nu}\hat{v}_{\ib}^\nu\Big).
 \eea
Remembering that the magnetic field is divergence free, we can
compare the first two terms of the above equation. Their ratio in
Fourier space is of order $(\tau_{\rm evo}\eta_{\rm C,eff}
k^2)^{-1}\simeq \tau_{\rm evo}/\eta_{\rm C,eff}$. Therefore, on
all scales of interest, we can conclude that the contribution of
the $\eta_{\rm C,eff}\,{\rm curl}\,  \hat{B}^\mu$ term in
(\ref{mezzobis}) is negligible compared to the last term.

The above considerations remain valid once we approach
recombination time, as long as the residual fraction of free
electrons $x_\ie$ is not too small. This is to ensure that the
approximations of the ratios of time scales made to obtain
(\ref{mezzobis}) remain valid. This is indeed the case, and it can
be checked from Fig.~\ref{fig:times}, since  $x_\ie \sim
10^{-3}-10^{-4}$ after last scattering
\cite{Seager1999a,Takahashi:2008gn} until reionization.

We are therefore left with the following expression for the
electric field produced by the tiny velocity difference between
electrons and protons:
 \bea
\label{GenerationE} E^\mu &=& -\frac{1-\beta^3}{1+\beta}
\frac{4\rho_\gamma \st}{3 e}\Big(\Delta v_{\ib\gamma}^\mu
+\frac{2}{5}\Theta^{\mu}_{\nu}v_{\ib}^\nu\Big).
 \eea
It is important to note that this expression does not contain the
number density of free electrons $n_\ie$. Therefore the electric field
produced by this mechanism before recombination is still present
after last scattering (see also \cite{Takahashi:2008gn}) and can
in principle continue to generate a magnetic field after
recombination.

We can now also finally prove that
 \be
\Delta v_{\ip \ie}^\mu \ll \Delta v_{\ie\gamma}^\mu\,.
 \ee
Using (\ref{GenerationE}) and (\ref{EqBroughly}) without the
$\eta_{\rm C,eff}$ term, and in the Maxwell equation
(\ref{Eqdvep}) leads to an estimation of the order of magnitude of
velocity differences:
 \bea\label{Goodlimitvelocities}
\Delta v_{\ip \ie}^\mu \propto \frac{\etaC\tau_{\rm C}} {\tau_{\rm evo} \tau_{\rm T}} \Delta v_{\ib \gamma}^\mu\,,~~ \Delta
v_{\ip \gamma}^\mu \simeq \Delta v_{\ib \gamma}^\mu \simeq \Delta
v_{\ie \gamma}^\mu \,.
 \eea
The order of magnitude of the ratio $\Delta v_{\ip \ie}^\mu/\Delta
v_{\ib \gamma}^\mu$ is shown in Fig.~\ref{fig:times} and remains
well below unity for all relevant times, even when Coulomb
scattering becomes less efficient than Compton scattering, that is
for $z \gtrsim 10^6$.

It also follows from (\ref{Goodlimitvelocities}) and (\ref{Collterms2}) that we can
rewrite (\ref{GenerationE}) as
\be
\label{GenerationE2} e(n_\ie+n_{\rm H})x_\ie E^\mu =
C^{\mu\perp}_{\ib\gamma}= \nabla_\nu T_{\ib}^{\mu\nu \perp},
 \ee
where we neglect terms of order $\beta$ and where here the baryon
index $\ib$ encompasses protons, electrons and hydrogen atoms.

As a conclusion of this section, we stress again that when we
assume that electrons and photons are tightly coupled, as was
originally considered in \cite{1970MNRAS.147..279H}, then the
electrons and protons are even more tightly coupled by the
electromagnetic field which is generated, so that the electrons
and protons can still be considered, from the point of view of
photons, as a single fluid of baryons. As a consequence, taking
$\Delta v_{\ie\gamma }^\mu \to 0$ at early times has to be
performed consistently by keeping $\Delta v_{\ip \ie}^\mu \ll
\Delta v_{\ie\gamma }^\mu$ when taking the limit. For the
tight-coupled limit, this is crucial, since it corresponds exactly
to the limit $v_{\ie}=v_{\gamma}=v_{\ip}=0$, and the collision
terms cannot be evaluated directly from their expressions
(\ref{Collterms1})--(\ref{Collterms4}).

\subsection{Local inertial frame (tetrad)}

It is convenient to express all quantities in a local inertial
frame, defined by an orthonormal tetrad $\bm{e}_{\aT}~
(\aT=0,1,2,3)$:
 \be
\label{deftetrad} e_{\aT}{}^{\mu} e_{\bT}{}^{\nu} g_{\mu
\nu}=\eta_{\aT \bT},\quad e^{\aT}{}_{\mu} e^{\bT}{}_{\nu} g^{\mu
\nu}=\eta^{\aT \bT}.
 \ee
The  tetrad indices are distinguished from general coordinate
indices by underlining, and $\iT,\jT,\kT\dots=1,2,3$. We choose a comoving tetrad, so that
$\bm{ e}_{\zT}$ is the fundamental observer 4-velocity:
$e_{\zT}{}^{\mu}=u^\mu$. In the background, $\bar{e}_{\zT}{}^{\mu}
= \bar{u}^\mu=(a^{-1},0)$. The perturbed tetrad is given in
Appendix~\ref{tetrads}. Derivatives along the tetrad vectors are
defined by
 \be\label{Defderiveetetrade}
\partial_\aT \equiv e_\aT{}^{\bB}\partial_\bB\,.
 \ee
Covariant derivatives in the tetrad frame are computed using the
affine connections given in Appendix~\ref{tetrads}.

Tetrads make the physical meaning of all nonscalar quantities more
transparent. In linear perturbation theory, it is common practice
to decompose perturbed quantities in a background tetrad. For
instance the velocity is often decomposed as $u^{\iB}_{(1)}\equiv
a^{-1 }v^{\iB}_{(1)}$, together with $u^{(1)}_{\iB}=a
v^{(1)}_{\iB}$, which means implicitly that $v^{(1)}_{\iB}\equiv
\delta_{\iB \jB}v^{\jB}_{(1)}$. Thus $v^{\iB}_{(1)}$ coincides
with $v^{\iT}_{(1)}=\bar {e}^{\iT}{}_{\jB}u^{\jB}_{(1)}$.
Introducing tetrads is the natural generalization of this standard
procedure when considering higher order perturbations, and this
has already been used for example to decompose velocities
\cite{Senatore:2008vi,Fitzpatrick:2009ci}. The nonlinear evolution
of the distribution of photons is well suited to computation in a
tetrad frame \cite{Pitrou2008}.

\subsection{Magnetic field}

The Maxwell equation (\ref{MaxwellBdot}) becomes in the tetrad
basis
 \be \label{MaxwellBdot1}
\partial_{\zT}(a^2 B^\iT)=-a^2 \epsilon^{\iT\lT\kT}
\partial_{\lT}\left[\left(1+\Phi-\Psi \right)E_\kT \right],
 \ee
Equivalently we can use derivatives in the coordinate basis:
 \be\label{Maxwell}
\left(a^2 B^\iT\right)'=-a^2
\epsilon^{\iT\lT\kT}\partial_{\ell}\left[\left(1+\Phi-\Psi
\right)E_\kT \right],
 \ee
where we have used the fact that the electric field is at least a
first order quantity, and the magnetic field a second order
quantity. The gravitational potentials in this expression occur
only at first order. Equation (\ref{Maxwell}) is compatible with
\cite{Maeda2008}, which can be seen via $E_\kT =
e_\kT{}^{\iB}E_{\iB}$.

To obtain (\ref{MaxwellBdot1}), we need
 \be
({\rm curl}\, E)^\iT = \epsilon^{\iT \lT \kT} \nabla_\lT E_\kT=
\epsilon^{\iT \lT \kT}
\partial_\ell \left[(1-\Psi) E_{\kT}\right],
 \ee
which uses the affine connections up to first order given in
Appendix~\ref{tetrads}. Also,
 \be
e^\iT{}_\mu\epsilon^{\mu\nu\lambda}\dot{u}_\nu E_\lambda
=\epsilon^{\iT\lT\kT}\,\dot{u}_\lT E_\kT = -\epsilon^{\iT\lT\kT}
E_\lT \partial_\kT \Phi  \,,
 \ee
which follows from
 \be
\dot{u}_\iT= (u^\mu\nabla_\mu u_\nu) e_\iT{}^{\nu}=(\nabla_{\zT}
e^{\zT}{}_\nu) e_\iT{}^\nu =
\Omega_{{\zT}\phantom{{\zT}}\iT}^{\phantom{{\zT}}{\zT}}
=\partial_\iT \Phi\,.
 \ee
In addition, we omitted terms like $\Phi \epsilon^{\iT \lT
\kT}\partial_\lT E_\kT$ and $\Psi\epsilon^{\iT \lT \kT}
\partial_\lT E_\kT$ in deriving (\ref{Maxwell}),
since the electric field contributes only at
first order -- and at this order, it is curl-free. For the same
reason, we can also replace $\partial_\lT$ by
$a^{-1}\partial_{\ell}$.

In summary, magnetogenesis is governed by
(\ref{Maxwell}) and (\ref{GenerationE2}), i.e.
 \bea
\hspace*{-1.5em}
&&\left(a^2 B_\iT\right)'=  -{a^2 \over e(  n_\ie+   n_{\rm H})x_\ie}
\epsilon_{\iT\lT\kT}\partial^{\ell}\left[\left(1+\Phi-\Psi \right)
C^{\kT}_{\ib\gamma} \right] \nonumber\\
&&~~=  -{a^2 \over e(  n_\ie+  n_{\rm H})x_\ie}
\epsilon_{\iT\lT\kT}\partial^{\ell}\left[\left(1+\Phi-\Psi \right)
\nabla_\nu T_{\ib}^{\nu\kT} \right], \label{SimpleunderstandingB}
 \eea
where here, as in (\ref{GenerationE2}), the baryon index $\ib$ encompasses
electrons, protons and hydrogen atoms.
Finally, note that the value of the magnetic field depends of
course on the observer. Its value in the baryon frame is related
to its value (\ref{Maxwell}) in the fundamental frame by
 \be\label{BChgFrame}
B^\iT_\ib=B^\iT-\epsilon^{\iT \lT \kT}\,v_{\ib\,\lT} E_\kT\,.
 \ee

\subsection{Numerical computation}\label{SecNumComp}

In order to solve the evolution equation for the magnetic field,
we need to solve the Boltzmann hierarchy for baryons and photons,
to compute the source of the electric field in
(\ref{GenerationE}). The basic idea is to decompose the
directional dependence of radiation in the local inertial frame
into multipoles:
 \bea \hspace*{-1em}
\Theta_{\iT_1\!\cdot\cdot{\iT_\ell}}(\bm{x})n^{\iT_1}\!\cdot\cdot
n^{{\iT_\ell}}\!\!&=&\!\!\!\!\int\!\!\frac{\dd^3 \gr{k}}{(2
\pi)^{3/2}}\! \sum_m\!
\Theta_\ell^m\!(\gr{k}) G_{\ell m}(\gr{k},\bm{x},\bm{n}) \\
G_{\ell m}(\gr{k},\bm{x},\bm{n})&=&\ii^{-\ell} \Big({4\pi \over
2\ell+1 }\Big)^{1/2} {\rm e}^{\ii k_\iB x^\iB} Y^{\ell m}(n^\iT)
\,.\label{DefGlm}
 \eea
We suppress the time dependence for convenience.

Terms quadratic in first order perturbations appear as
convolutions, and we introduce the notation
 \be
\mathcal{K}\{f_1f_2\}(\bm{k}) \equiv \int \frac{\dd^3\gr{k}_1
\dd^3\gr{k}_2}{(2 \pi)^{3/2}}\, \delta_{\rm
D}^{3}(\gr{k}_1+\gr{k}_2-\gr{k})f_1(\bm{k}_1) f_2(\bm{k}_2)\,.
 \ee
A Fourier mode $q_\iB$ is decomposed on the helicity basis of the
background spacetime as
 \bea
q^{\iB} &=& \delta^{\iB \jB }q_{\jB}= q_{(+)} \bar{
e}_{(+)}^{\iB}+ q_{(-)} \bar{ e}_{(-)}^{\iB} + q_{(0)}
\bar{e}_{(0)}^{\iB}\,,\\ q_{(h)} &=& q_{\iB} \bar{
e}_{(h)}^{*\,\iB}\,.
 \eea
The background helicity basis vectors $\bar{\bm{e}}_{(h)}$, with
helicity $h=0,\pm$   are defined in \cite{Pitrou2008}. The
azimuthal direction $h=0$ corresponds to scalar perturbations and
is aligned with the total Fourier mode, { i.e.} $\bar{\bm{
e}}_{(0)} = \hat{\bm{ k}} $, while $h=\pm$ correspond to vector
perturbations. At first order, when the mode is aligned with the
azimuthal direction since ${\bm q}={\bm k}$, there are only scalar
perturbations. For vector quantities like the electric field, we
need to use a helicity basis $\bm{e}_{(h)}$ on the perturbed spacetime, and this is built by the identification of $\bar{\bm{
e}}_{(h)}$ with $\bm{ e}_{(h)}$, i.e. $\bar{e}_{(h)}^{\iB} =
e_{(h)}^\iT$. Vector quantities like the electric field $E_\iT$
are then expanded as
 \bea
X^{\iT}& =& X_{(+)} { e}_{(+)}^{\iT}+ X_{(-)} { e}_{(-)}^{\iT} +
X_{(0)} {e}_{(0)}^{\iT}\,,\\ X_{(h)}& =& X_\iT { e}^{*
\iT}_{(h)}\,.
 \eea

In this basis, the Maxwell equation (\ref{Maxwell}) becomes
(explicitly giving the perturbative order of quantities)
 \bea \hspace*{-2em}
&&\Big[a^2 B_{(\pm)}^{(2)}(\bm{k})\Big]'= \nonumber \\
\hspace*{-2em} &&~~~~~ \mp k a^2
\Big[E_{(\pm)}^{(2)}(\bm{k})+{\cal
K}\Big\{\big[\Phi^{(1)}-\Psi^{(1)}\big]
E_{(\pm)}^{(1)}\Big\}(\bm{k}) \Big].
 \eea
We projected (\ref{Maxwell}) along $e^{(h)*}_\iT$ and used
 \be
\ii\epsilon^{\iT\lT\kT} k_\lT e^{(\pm)}_{\kT}=\pm k
e_{(\pm)}^\iT\,,~~ \ii e^{(\pm)*}_\iT\epsilon^{\iT\lT\kT} k_\lT
X_{\kT}=\pm k X_{(\pm)}\,.
 \ee
Note that there are only contributions from $h=\pm $ and we thus
recover that scalar perturbations cannot generate a magnetic field
and vortical perturbations are required to source the magnetic
field. Using the multipole decomposition of (\ref{GenerationE}),
and neglecting $\beta\ll 1$, we obtain finally,
  \bea
\label{MFeq1} &&\hspace{-2em} \Big[a^2 B_{(\pm)}^{(2)}(\bm{
k})\Big]'= \pm k a^2 \frac{4 \st\bar \rho_{\gamma}}{3
e}\Big[V_{(\pm)}^{(2)}(\bm{ k})
\nonumber   \\
&&+{\cal K}\Big\{\big[\delta_{\gamma}^{(1)}
+\Phi^{(1)}-\Psi^{(1)}\big] V_{(\pm)}^{(1)}\Big\}(\bm{k})
\nonumber \\
&&-{\cal K}\Big\{\sum_{h}\frac{\kappa({\pm1},h)}{5}
\Theta^{\pm1+h(1)}_2 v_{\ib(-h)}^{(1)}\Big\}(\bm{k})
\Big]\nonumber\\
&&\equiv \pm k a^2 \frac{4 \st\bar \rho_{\gamma}}{3e}
\left[S^{(\pm)}_1(\bm{ k})+S^{(\pm)}_2(\bm{ k})+S^{(\pm)}_3(\bm{
k})\right],
 \eea
where
 \be
V_{(h)} \equiv v_{\ib (h)} -v_{\gamma (h)}\,,
 \ee
and $\delta_\gamma= \delta \rho_\gamma/\bar \rho_\gamma$. Also,
 \bea
\hspace{-0.1cm}\kappa(h,0) = \sqrt{(4-h^2)},\, \kappa(h,\pm1) =
-\sqrt{\frac{(2\pm h)(3 \pm h)}{2}}\!.
 \eea
The last equality in (\ref{MFeq1}) defines the contribution of
each line above: $S^{(\pm)}_1$ is the purely second order
contribution from $V^{(2)}$; $S^{(\pm)}_2$ is the $\delta_\gamma
V$ contribution and $S^{(\pm)}_3$ is the $\Theta_2v_\ib$
contribution.

Although $V^{(1)}_{(\pm)}(k)$ vanishes at first order since there
are no vector perturbations, $V^{(1)}_{(\pm)}(\gr{k}_1)$ and
$V^{(1)}_{(\pm)}(\gr{k}_2)$ do not vanish in general, since the
modes $\gr{k}_1$ and $\gr{k}_2$ are not necessarily aligned with
the azimuthal direction $\hat{\gr{k}}= \gr{k}/k$. We first need to
obtain their expression when the modes $\gr{k}_1$ or $\gr{k}_2$
are aligned with the azimuthal direction, and then we perform a
rotation of the azimuthal direction \cite{Pitrou2008}.

In order to explicitly take into account the symmetry of the
convolution products in (\ref{MFeq1}), we can symmetrize the
source terms. At first order there are only scalar perturbations,
and all first order tensorial quantities are gradients of scalar
functions, so that
$X^{(1)}_{\iT_1\dots\iT_n}=X^{(1)}_{\iB_1\dots\iB_n}=
\partial_{\iB_1}\dots\partial_{\iB_n}X^{(1)}$. Most of the source
terms are of the form $\epsilon^{\iT\lT\kT}\partial_\ell \left(X
\,Y_\kB\right)=\epsilon^{\iT\lT\kT}\partial_\ell \left(X
\partial_\kB Y\right)$, and once projected along
$e^{(\pm)*}_\iT$ they contribute to the generation of the magnetic
field proportionally to
 \bea \hspace*{-2em}
&&\bar{e}_{(\pm)}^{*\,\,\iB} [X \partial_\iB Y ](\gr{k})=
\nonumber\\ \hspace*{-2em}&&\frac{\ii}{2}\!\int\! \frac{\dd^3 \bm{
q}}{(2 \pi)^{3/2}} q_{(\pm)}\left[X(\bm{k-q})Y(\bm{ q})-X(\bm{
q})Y(\bm{ k-q})\right] \!.
 \eea
Here $X$ and $Y$ denote $\delta_\gamma, V^{(1)}, v_{\rm b},\Phi,
\Psi$.

This symmetrization, which is always possible, shows that for
these types of terms, the configurations of
$(\gr{k},\gr{k}_1,\gr{k}_2)$ with $k_1=k_2$ will not contribute in the
convolution. Only couplings from a quadrupolar quantity to
gradient terms, which are of the type
 \be
\epsilon^{\iT\lT\kT}\partial_\ell \left(X_\kB^{j}
\partial_j Y\right)=\epsilon^{\iT\lT\kT}\partial_\ell
\left(\partial_\kB \partial^j X \partial_j Y\right)\,,
 \ee
as in the last line of (\ref{MFeq1}), can have contributions to
the convolution coming from configurations with $k_1=k_2$. The
generated magnetic field is thus severely suppressed at early
times for these configurations since the quadrupole of radiation
is suppressed in the tight-coupling regime.

\section{Numerical results}\label{results}

\subsection{Transfer functions}

In order to obtain the final magnetic field spectrum produced via
this mechanism, we integrate numerically the evolution equations
for cosmological perturbations up to second order, since we have
to take into account even the behavior of the second order
velocity difference between baryons and photons $V^{(2)}_{(h)}
(k,\eta)$. We use throughout the cosmological parameters of WMAP7
\cite{Komatsu:2010fb}.

For a variable $X$, the first order transfer function is
$X^{(1)}(\gr{k},\eta)={\cal X}^{(1)}(k,\eta)\Phi_{\rm
in}(\gr{k})$, where $\Phi_{\rm in}$ is the gravitational potential
deep in the radiation era. Because of statistical isotropy, the
first order transfer function depends only on the magnitude of the
Fourier mode and not on its direction. This is however only
strictly true for multipoles like $\Theta_2^m$ and $V_{(h)}$
defined from non-scalar quantities if the azimuthal direction is
aligned with $\hat{\gr{k}}$, and considering only scalar
perturbations at first order the contributions for $h\neq 0$
vanish. However, when using these first order transfer functions
in the quadratic terms of the second order equations, we must
rotate these multipoles according to the angles between
$\hat{\gr{k}}_1,\hat{\gr{k}}_2$ and $\hat{\gr{k}}$. This is to
ensure that the multipoles remain defined with respect to the
total momentum $\hat{\gr{k}}$ \cite{Pitrou2008}.

The second order transfer function ${\cal X}^{(2)}
(\gr{k}_1,\gr{k}_2,\eta)$ is defined by
 \be
X^{(2)}(\bm{ k},\eta)={\cal K}\left\{{\cal
X}^{(2)}(\gr{k}_1,\gr{k}_2,\eta) \Phi_{\rm in}(\gr{k}_1)\Phi_{\rm
in}(\gr{k}_2)\right\}(\bm{k}).
 \ee
Without loss of generality we enforce ${\cal
X}^{(2)}(\gr{k}_1,\gr{k}_2,\eta)={\cal
X}^{(2)}(\gr{k}_2,\gr{k}_1,\eta)$ in numerical calculations. The
transfer functions of the first and second order quantities needed
in the source terms
are obtained by a joint solution of the Boltzmann equation (for
photons and neutrinos), the conservation and Euler equations (for
baryons and cold dark matter) and the Einstein equations (for
metric perturbations). They are found numerically using the same
techniques as in \cite{2010JCAP...07..003P}.

The transfer function of the magnetic field can be split into the
different contributions of the $S_i^{(\pm)}$ sources defined in
(\ref{MFeq1}). The transfer functions of these contributions are
related to the transfer functions of the sources through
 \be\label{EqBfroms}
{\cal B}^{S_i}_{(\pm)}(\gr{k}_1,\gr{k}_2,\eta)= \frac{ 4\st k}{3e
a^2} \int^\eta \dd \eta' a ^2 \bar \rho_\gamma {\cal
S}^{(\pm)}_i(\gr{k}_1,\gr{k}_2,\eta')\,,
 \ee
and this is how we obtain the complete time behavior of the
magnetic field. A crucial point that will turn out to have
important consequences is that the final redshift for numerical
integration should be taken after the recombination epoch. The
electric field that results from the small electron-proton
velocity difference and that gives rise to a magnetic field is
still present after last scattering, when the fraction of free
electrons $x_\ie$ is tiny but still does not completely vanish
(see also \cite{Takahashi:2008gn}).

In order to compute the equal time correlation functions of the
magnetic field, we need the power spectrum of the initial
potential, defined by
 \be
\langle\Phi_{\rm in}(\gr{k})\Phi_{\rm in}^*(\gr{q}) \rangle \equiv
\delta(\gr{k}-\gr{q})P(k).
 \ee
If the source terms are Gaussian random variables, we can apply
Wick's theorem, and the contributions of the two polarizations
$h=\pm$ add up quadratically:
 \bea
&&\hspace{-0.45cm}\langle \bm{ B}(\bm{ k},\eta)\bm{ B}^*(\bm{ k'},\eta)\rangle
\nonumber\\
&&\hspace{-0.45cm}=\frac{2\delta^3_{\rm D}(\bm{ k}-\bm{ k'})}{(2\pi)^3} \int
\dd^3\gr{q}\,
        P(q)P(|\bm{ k-q}|) \times \nonumber \\
&&\hspace{-0.45cm}\Big\{|{\cal B}_{(+)}(\gr{q},\gr{k}-\gr{q},\eta)|^2+ {\cal
B}_{(+)}(\gr{q},\gr{k}-\gr{q},\eta){\cal B}^*_{(+)}(\gr{k}-\gr{q},
    \gr{q},\eta) \Big\}\nonumber\\
&&\hspace{-0.45cm}=\frac{4\delta^3_{\rm D}(\bm{ k}-\bm{ k'})}{(2\pi)^3} \int \dd^3\gr{q}\,
P(q)P(|\bm{ k-q}|)|{\cal B}_{(+)}(\gr{q},\gr{k}-\gr{q},\eta)|^2\nonumber\\
&&\hspace{-0.45cm}\equiv \delta^3_{\rm D}(\bm{ k}-\bm{ k'})
P_B(k,\eta),\label{Bspectrum}
 \eea
where ${\cal B}_{(\pm)} = \sum_i {\cal B}_{(\pm)}^{S_i}$. In the
last line we have defined the power spectrum of the magnetic field
$P_B$. Its value today is plotted in Fig.~\ref{fig1}.

In order to have a deeper analytical understanding of the
resulting magnetic field spectrum, we study each contribution
$S_i$ independently. There are cross correlations in
(\ref{Bspectrum}), but our aim is to assess the relative
importance of the different contributions; the $P_B^{S_i}$ are
defined by replacing ${\cal B}_{(+)}$ with ${\cal B}_{(+)}^{S_i}$
in (\ref{Bspectrum}).

\subsection{$\delta_\gamma \Delta v_{\ib \gamma}$ contribution}

\begin{figure*}[!htp]
\includegraphics[width=8.5cm]{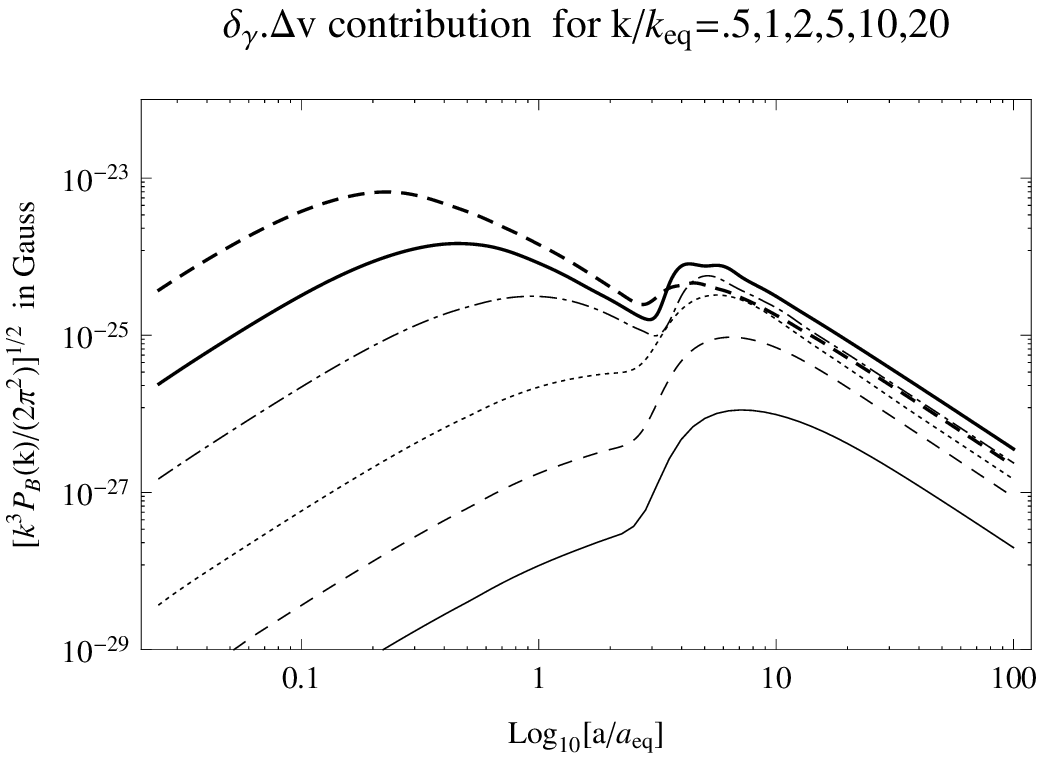} \quad
\includegraphics[width=8.5cm]{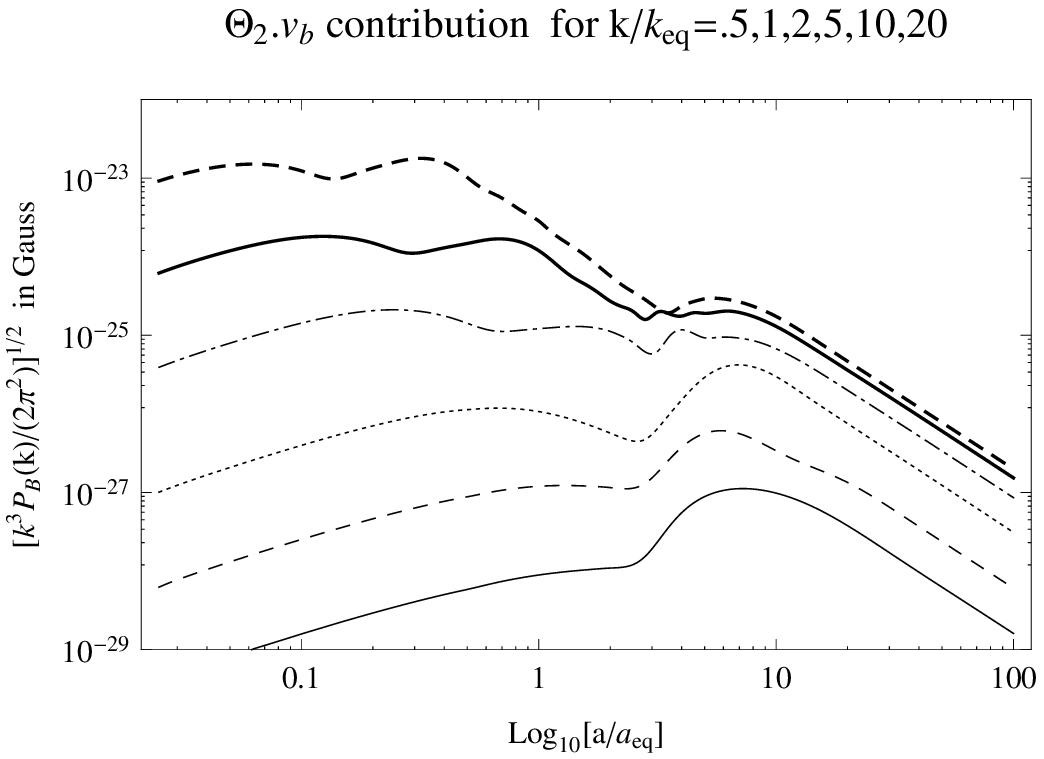}
\caption{{\em Left:} Magnetic field spectrum $P_B^{S_2}(k, \eta)$
from only the $S_2$ contribution in (\ref{MFeq1}), for different
$k/k_{\rm eq}$, with values increasing from bottom to top. {\em
Right:} Magnetic field spectrum $P_B^{S_3}(k, \eta)$ from only the
$S_3$ contribution in (\ref{MFeq1}).} \label{S2}
\end{figure*}

The velocity difference between baryons and photons is severely
suppressed in the tight-coupling limit relative to other
perturbations like $\delta_\gamma$; we expand this tiny velocity
difference in terms of the expansion parameter $k/\tau' \ll1$,
where $\tau'=n_\ie  \st a$ is the derivative of the
optical depth for Thomson scattering. At first order in $k/\tau'
$, in the radiation-dominated background on super-Hubble scales,
 \bea
V_{(0)}^{(1)}(k,\eta)
&\simeq & R \frac{k}{\tau'}\left(\frac{\delta_\gamma}{4}-
\frac{\mathcal{H} v_{\ib (0)}}{k}\right)\propto k^3
\frac{\eta^5}{\eta_\eq^2}\,,\\
\delta_\gamma(k,\eta) &\simeq& {\rm const}\,.
 \eea
Using $R=3\bar \rho_\ib/(4\bar \rho_\gamma)\propto a$, $1/\tau'
\propto a^{-2}$ and $a\propto \eta$, we get
 \bea
{\cal S}^{(+)}_2(|\bm{ k-q}|,q,\eta) \propto \hat q_{(+)}
\left(q^3-|\gr{k}-\gr{q}|^3\right) \frac{\eta^5}{\eta_\eq^2}~.
 \eea
Then (\ref{EqBfroms}) gives the early-time and large-scale
behaviour of ${\cal B}_{(\pm)}^{S_2}$, and the resulting magnetic
field power spectrum behaves as
 \bea\label{EqconvolforPB}
P^{S_2}_B(k,\eta)&\propto& k^2  \int \dd^3q\, |\hat q_{(+)}|^2
P(q)P(|\bm{ k-q}|)  \nonumber \\
&& \quad \times \left[ q^6 - q^3 |\bm{ k-q}|^3 \right]
\frac{\eta^4}{\eta_\eq^4}\,.
 \eea
For a scale-invariant initial power spectrum, $P(q) \propto
q^{-3}$,
 \be
\label{EqScalingSdek} P^{S_2}_B(\lambda k,\eta) = \lambda^5
P^{S_2}_B(k,\eta),
 \ee
as can be seen just by a change of variable in the integral of
(\ref{EqconvolforPB}). In \cite{Ichiki:2007hu} it is found that
$P^{S_2}_B(\lambda k,\eta) = \lambda^4 P^{S_2}_B(k,\eta)$. The disagreement appears to arise since \cite{Ichiki:2007hu} infer the
dependence on $k$ from the $q\gg k$ contribution to the integral
in (\ref{EqconvolforPB}) -- but the main contribution to that
integral are also limited to $q \lesssim k$ given the argument at the
end of section \ref{SecNumComp}. We finally find that for the
$S_2$ source term, the power spectrum of the magnetic field
behaves as
 \be
\sqrt{k^3 P_{B}^{S_2}(k,\eta)}\propto k^4
\frac{\eta^2}{\eta_\eq^2}\,. \label{s2magpow}
 \ee
This behaviour in $k$ and $\eta$ at early times when the mode is
still super-Hubble, is confirmed by numerical integration, as is
evident from Fig.~\ref{S2} (left).

\subsection{$\Theta_2 v_{\ib}$ contribution}

Similar analytical arguments apply to the magnetic field generated
by the source $S_3$. The tight coupling expansion of the source is
 \be
\Theta^0_2(k,\eta)\propto \frac{k}{\tau'} v^\gamma_{0} \propto k^2
\frac{\eta^3}{\eta_\eq},~~ v_{\ib(0)}(k,\eta) \propto k\eta,
 \ee
in a radiation background on super-Hubble scales. This implies
that the $S_3$ contribution to the magnetic field power spectrum
behaves as
 \be \label{s3magpow}
\sqrt{k^3 P_{B}^{S_3}(k,\eta)}\propto k^4 \frac{\eta}{\eta_\eq}\,.
 \ee
It has the same $k$ dependence as (\ref{s2magpow}) but a different
$\eta$ dependence. The analytical form is verified by the
numerical output shown in Fig.~\ref{S2} (right).

\subsection{$\Delta v_{\ib \gamma}^{(2)}$ contribution}

\begin{figure*}[!htp]
\includegraphics[width=8.5cm]{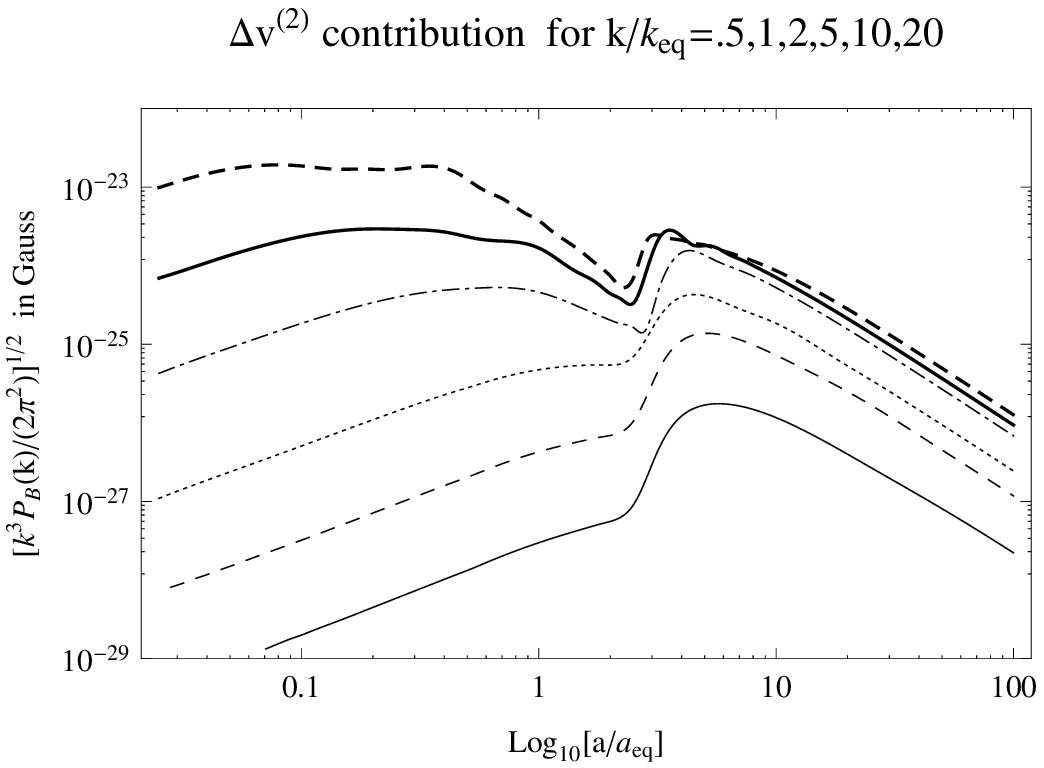}\quad
\includegraphics[width=8.5cm]{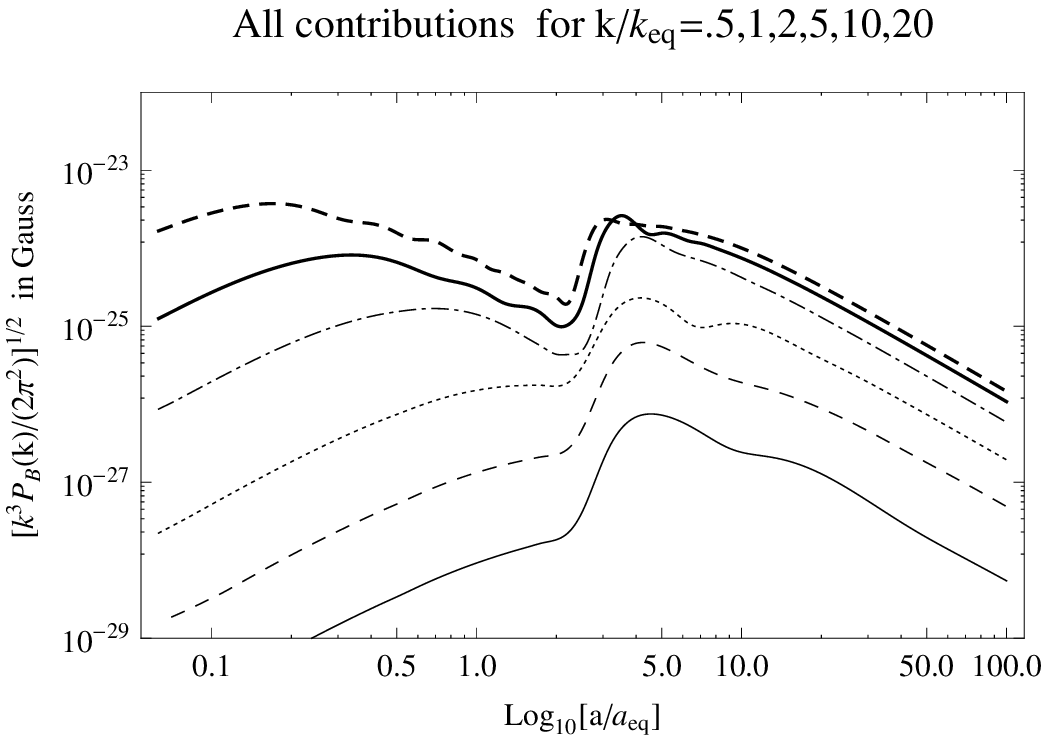}
\caption{{\em Left:} Magnetic field spectrum $P_B^{S_1}(k, \eta)$
from only the $S_1$ contribution in (\ref{MFeq1}), for different
$k/k_{\rm eq}$, with values increasing from bottom to top. {\em
Right:} Magnetic field spectrum $P_B(k, \eta)$ for all
contributions.} \label{S4}
\end{figure*}

For the purely second order part $S_1$, the only way to assess its
contribution is to consider the tight coupling expansion of the
evolution equation for the vorticity of baryons. Indeed, we need
to evaluate first the total contribution $\sum_{i}{\cal S}_i$ at
lowest order in tight-coupling, and the detail of this derivation
is given in Appendix~\ref{AppendixVorticity}. It follows that
$\sum_i {\cal S}_i$ behaves as $(k/\tau') (k \eta)^2 \propto k^3
\eta^5/\eta^2_\eq$, which implies that for the total magnetic
field
 \be
\sqrt{k^3 P_{B}(k,\eta)}\propto k^4 \frac{\eta^2}{\eta_\eq^2}\,.
 \ee
This behaviour is confirmed by numerical integration, as shown in
Fig.~\ref{S4} (right).
Since $ {\cal S}_2\propto k^3 \eta^5/\eta^2_\eq$, ${\cal S}_3
\propto k^3 \eta^4/\eta_\eq$, and $ \sum_i {\cal S}_i \propto k^3
\eta^5/\eta^2_\eq$, we obtain that ${\cal S}_1 \propto k^3
\eta^4/\eta_\eq$.
Thus $S_3$ contributes to the magnetic field power spectrum as
 \be
\sqrt{k^3 P_{B}^{S_1}(k,\eta)}\propto k^4 {\eta \over \eta_\eq}\,,
 \ee
which is verified in Fig.~\ref{S4} (left).

\subsection{Magnetic power spectrum}

From these plots it is evident that the magnetic field is still
generated after recombination. This is the reason that it is
important, to set the final time of integration after
recombination, since the largest contribution comes from this last
period of generation. Indeed, before reaching the usual `final'
stage where the magnetic field is no longer sourced but only
redshifts with time ($B\propto a^{-2}$), we observe a bump in the
resulting magnetic field spectrum, corresponding to the
recombination time. This should be interpreted as an increase in
magnetic field generation due to decoupling of photons and
baryons.

In the decoupling regime the fluid of photons and baryons is no
longer equivalent to a perfect fluid. The departure from tight
coupling may be interpreted via non-adiabatic pressure
perturbations, which can source the total vorticity
\cite{Kobayashi:2007wd,Lu:2008ju,Christopherson:2009bt,Christopherson:2010dw}.
It is not a priori evident that this could lead to an increase in
the magnetic field generation. On the one hand, the total
vorticity is sourced when interactions between baryons and photons
are less efficient, but on the other hand, there is less vorticity
exchange between photons and baryons since the collisions are less
efficient. In the ideal limit where the decoupling is complete,
the vorticity of photons and baryons is adiabatically evolving
according to (\ref{Eqadiabaticvorticity}), whereas the total
vorticity is sourced by the gradients in the total non-adiabatic
pressure. This is possible because the vorticities of the
different fluids do not add up linearly to give the total
vorticity as can been seen from (\ref{EqExpressionomega}).

However, when decoupling occurs, we observe that there is in fact
an increased generation of magnetic field in that phase, and this
essentially comes from the factor $x_\ie$ in
(\ref{SimpleunderstandingB}), i.e. from the fact that the magnetic
field is generated via the residual ionized fraction. More
precisely, the generation of the magnetic field is proportional to
$\partial_{[\jB}\nabla_\mu T^{\mu}_{\ib\,\,\kT]}/x_\ie$ and not
only to $\partial_{[\jB}\nabla_\mu T^{\mu}_{\ib\,\,\kT]}$, so even
when $\nabla_\mu T^{\mu}_{\ib\,\,\kT}\to 0$ around decoupling,
$\nabla_\mu T^{\mu}_{\ib\,\,\kT}/x_\ie$ can still have sizeable
values. This last significant stage of magnetic field generation
is counterbalanced and finally stopped by the redshifting of
photon energy density ($\bar \rho_\gamma\propto a^{-4}$). It can
be seen from (\ref{MFeq1}) that the background radiation energy
density controls the efficiency of the total magnetic field
production after recombination.

\begin{figure*}[!htp]
\includegraphics[width=8.5cm]{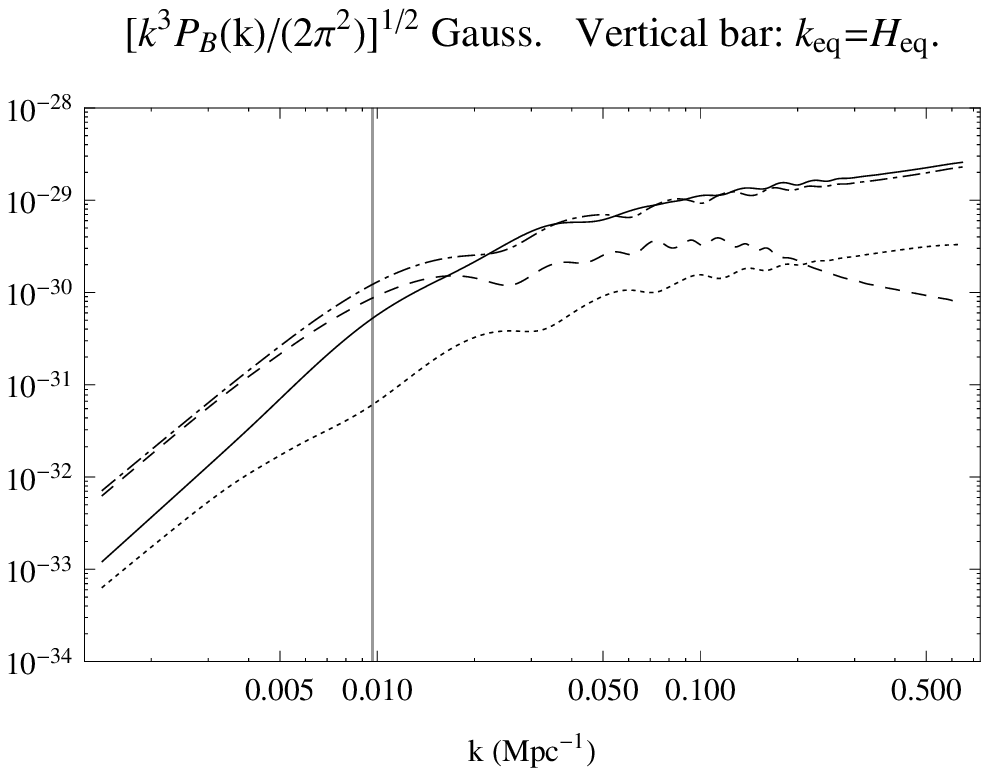}\quad
\includegraphics[width=8.5cm]{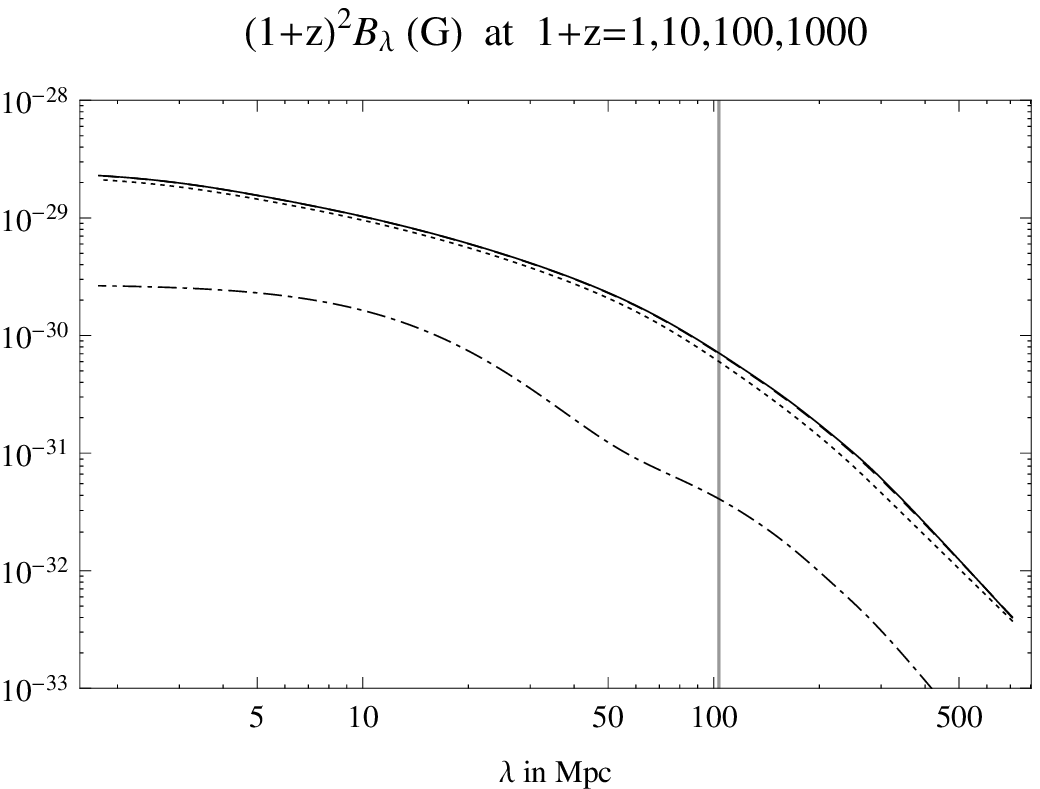}
\caption{{\em Left:} Magnetic field spectrum today (solid).
Contributions from the different sources in (\ref{b-sources}) are distinguished: second order velocity term $S_1$ (dot-dashed), quadratic term $S_2$ in velocity and density (dashed), quadratic term $S_3$ in anisotropic stress and velocity (dotted).
{\em Right:} Comoving magnetic field strength at a given scale at
times $1+z=1,10,100,1000$ corresponding respectively to solid,
dashed, dotted and dot-dashed lines. (Dashed and solid lines
cannot be distinguished).} \label{fig1}
\end{figure*}

The power spectrum of the magnetic field is shown in
Fig.~\ref{fig1} (left). The behaviour on large scales ($\propto k^4$) is explained
above. The behaviour on small scales is complex, since it depends
mainly on the generation between horizon crossing time and Silk
damping time. During that period, the analysis which we restricted
to super-Hubble scales does not apply -- and the adiabatic
redshifting does not apply either, since the magnetic field
continues to be generated.  For $k\gg k_{\rm eq}$, a reasonable linear approximation is $\log\left(\sqrt{k^3P_B}\right) \propto 0.5\log  k$.

\subsection{Magnetic amplitude}\label{SecBamplitude}

The magnetic field amplitude  smoothed over a comoving scale
$\lambda$ is
\bea
\label{BatscaleDefinition} B_{\lambda}^2 &=& \frac{1}{V} \int
\dd^3\bm{ y}  \langle  \bm{ B}(\bm{ x})
\bm{ B}^*(\bm{ x+y}) \rangle \exp\left( -\frac{y^2}{2\lambda^2}\right)
\nonumber \\
&=&\frac{1}{2\pi^2}\int_0^{k_{\rm damp}} \dd k\,k^2 P_B(k)
\exp\left(-\frac{k^2\lambda^2}{2}\right),
 \eea
where the normalization volume is $V=\int \dd^3\bm{ y} \exp[
-y^2/(2\lambda^2)]= \lambda^3 (2\pi)^{3/2}$.
Note that the integral is insensitive to the upper cutoff, which
may be taken to infinity, since $\lambda\gg \lambda_{\rm damp}$.
The magnetic field strength is shown in
Fig. \ref{fig1} (right).

The field strength at $10\,$Mpc is approximately $10^{-29}$ Gauss
and three times as much on cluster scales $1\,$Mpc. Given the
slope of the spectrum, this is expected to grow to larger values
for smaller scales. Our numerical integration does not allow us to
investigate smaller scales since the numerical integration time
increases dramatically with $k_{\rm max}$. In addition, the
results become unreliable on small scales where density
perturbations have become nonlinear by $z=0$. On the comoving scale of the Hubble radius at
equality, the strength is $\sim 10^{-30}\,$G.

\subsection{Frame dependence}

\begin{figure}[!htp]
\includegraphics[width=8.5cm]{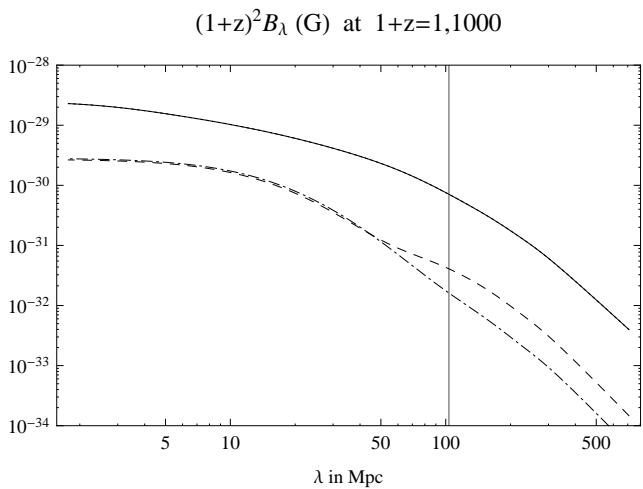}
\caption{Magnetic field strength at a given scale as measured in
the fundamental frame at $1+z=1$ (continuous) and $1+z=1000$
(dashed), and as measured in the baryon frame at $1+z=1$ (dotted)
and $1+z=1000$ (dot-dashed). Dotted and continuous lines cannot be
distinguished.} \label{FigBaryonsFrame}
\end{figure}

At early times when photons and baryons are tightly coupled, the
magnetic field measured in the baryon-photon fluid is not
generated at lowest order in the tight coupling expansion. This is
shown in Appendix~\ref{AppBevolutionBaryonsFrame}. Only higher
orders in the tight-coupling expansion contribute to
magnetogenesis. However, since most of the magnetic field
production occurs when the tight-coupling expansion breaks down
around recombination, this suppression is only relevant at early
times, before recombination, and for modes which remain for the
longest time in the tight-coupled regime, i.e. for large scales.
Therefore the difference between the magnetic field in the
fundamental frame and in the baryon frame decreases, and they are
nearly equal today, as shown in Fig.~\ref{FigBaryonsFrame}. This
shows that at $1+z=1000$ there is a suppression for large scales
in the baryon frame, but today there is no more suppression since
most of the magnetic field has been generated around recombination
time.

\section{Discussion and comparison with previous results}\label{SecComparison}

Our approach is the first complete analysis of magnetogenesis
around recombination, in the sense that it does not neglect any
term in the second order equation for the generation of the
magnetic field -- previous work has omitted at least one of the
terms. Therefore our results will necessarily differ from existing
partial results and we discuss briefly how some of these
differences arise.

Two general points can be highlighted:
\begin{itemize}\itemsep=-4pt
\item
Numerical computation is essential to obtain the magnetic power
spectrum -- and even for a reliable estimate of the magnetic field
strength. For example, \cite{Maeda:2008dv,Kobayashi:2007wd} use
similar analytical methods and incorporate the same source terms,
but the two estimated field strengths on the recombination Hubble
scale differ by orders of magnitude. A full numerical integration
is needed, especially to take into account all orders in the
tight-coupling expansion. This was initiated by
\cite{Ichiki:2007hu}, and we have built on their work.
\item
Neglecting any of the source terms for magnetogenesis not only
leads to inaccurate predictions -- it also misses the fact the
separate source terms do not simply add up linearly. The total of
the different contributions is suppressed in the tight-coupling
regime on large scales by a factor $(k\eta)^2$: the details are
given in Appendix \ref{magtcff}. As a consequence, discarding some
terms implies that this suppression in the tight-coupling regime
is neglected -- which leads to an overestimate of the magnetic
field generated. This is especially critical for the largest
scales where tight-coupling is valid at the latest times.
\end{itemize}

In \cite{Hogan:2000gv,Gopal:2004ut,Siegel:2006px} the anisotropic
stress contribution, $S_3$  in (\ref{MFeq1}), and the second-order
velocity contribution, $S_1$, are neglected. It is apparent from
the power spectrum plot in Fig. \ref{fig1} that both of these
contributions are substantial and cannot be neglected for a
reliable prediction of the magnetic field. In addition, these
references omit the scalar metric perturbations. Metric
perturbations and the second order velocity are included in
\cite{Matarrese2004,Kobayashi:2007wd,Maeda:2008dv}, but the
anisotropic stress is neglected.

In \cite{Ichiki:2007hu} the anisotropic stress is included, but
the second order velocity contribution is neglected.
In addition to this difference from our work, we find a different
time and momentum dependence for the large-scale and early-time
behaviour of the $S_2$ and $S_3$ contributions. We then find $\sqrt{k^3 P_B}
\propto k^4$ while they find $\propto k^{7/2}$.

The first 
numerical prediction of the magnetic power spectrum was given by
\cite{Matarrese2004}, neglecting anisotropic stress but including
second order velocity. However, our power spectrum is
significantly different from theirs. Part of the difference is due
to anisotropic stress, but there is a further difference arising
from the treatment of velocities. The evolution equation for the
magnetic field can be given by (\ref{SimpleunderstandingB}). It is
true that in the tight-coupled regime (see Appendix \ref{AppEuler}
for details), the velocities of electrons, protons and photons can
be approximated to be equal. However, it is erroneous to use
$\epsilon^{\iT\lT\kT}\partial_{\lB}\nabla_\mu
T^{\mu}_{\gamma\,\,\kT}=0$ to estimate the vorticity evolution.
Indeed, in order to cancel the collision term when taking the
tight-coupling limit, we have to consider a combination which uses
the action reaction law and for which the collision terms do not
appear. It is given by the total fluid vorticity conservation
equation:
 \be
\epsilon^{\iT\lT\kT}\partial_{\lB}\nabla_\mu
T^{\mu}_{\gamma\,\,\kT}+\epsilon^{\iT\lT\kT}\partial_{\lB}\nabla_\mu
T^{\mu}_{\ib\,\,\kT}=0\,.
 \ee
In the tight-coupled limit, the fluid of baryons and the fluid of
radiation exchange vorticity, essentially because the dilution of
their energy density is different, and this exchange of vorticity
is then required to maintain equal velocities at all times. In
\cite{Matarrese2004} it is implicitly assumed that $ C_{\gamma
\ie}^{\mu\perp}$ can be neglected because the velocity of
electrons is assumed to be close to that of photons. However, as
we discussed in Sec. \ref{SecMagneticField}, the limit has to be
consistent with (\ref{Goodlimitvelocities}), and this collision
term is precisely responsible for the vorticity exchange between
photons and electrons, and thus between photons and baryons -- and
it cannot be ignored. The vorticity evolution in the
tight-coupling limit should be computed using
(\ref{GenerationTermTC}), i.e. by substituting the tight-coupling
solution of velocities and energy densities perturbations in
(\ref{SimpleunderstandingB}).

In \cite{Kobayashi:2007wd} it is shown that there can be no
generation of magnetic field in the photon frame at strictly less
than the first order in tight coupling (if there is no initial
vorticity). Note that what we call first order in tight-coupling
(see also \cite{Pitrou:2010ai}) is called second order in tight
coupling by \cite{Kobayashi:2007wd,Maeda:2008dv}. In our case, we
focus on $C_{\ib\gamma}^\mu$, whereas they focus on $(k /\tau')
C_{\ib\gamma}^\mu $ where $\tau'$ is the interaction rate and
$k/\tau'$ is the parameter of the tight-coupling expansion. The
result of \cite{Kobayashi:2007wd} is compatible with our results
in Appendix \ref{AppBevolutionBaryonsFrame}, since in the
tight-coupled regime the photon frame is the baryon frame. Thus
the magnetic field in the photon frame will be generated only
starting from the next order, i.e. at first order in the
tight-coupling expansion. Our numerical approach does not rely on
a tight-coupling expansion since we integrate the full system of
equations, and in that sense we consider necessarily the full
tight-coupling expansion in our computation. We checked
numerically that at early times, when photon-baryon coupling is
efficient, the magnetic field in the baryon frame is severely
suppressed compared to the magnetic field in the fundamental
frame.

\section{Conclusion}

\label{conclusions}

We have performed for the first time a full numerical computation
of the seed magnetic field generated by nonlinear dynamics, taking
into account all general relativistic effects and all source terms. We discussed the range of applicability of
the mechanism on cosmological scales and concluded that the
generation of the magnetic field is directly related to the
Compton drag by photons on baryons. Even in the tight coupling
regime, photons exchange vorticity with baryons and the magnetic
field is created. Since the electric field that sources the
magnetic field does not depend on the fraction of free electrons,
the magnetic field is still generated after recombination, given
that there is a relic fraction of charged particles, and we find
that the largest production takes place in this final stage.

Our results are summarized in Fig. \ref{fig1one}. The power spectrum (left plot) behaves as
 \be
\sqrt{k^3 P_B} \propto \left\{ \begin{array}{ll} k^4 & k\ll k_{\rm eq} \\ k^{0.5} & k \gg k_{\rm eq} \end{array} \right.
 \ee
On cluster scales the
comoving field strength is (right plot)
 \be
B_{1\,{\rm Mpc}} \sim 3\times 10^{-29}\,{\rm G}.
 \ee

\acknowledgments

EF is supported by the Swiss National Science Foundation. CP is
supported by STFC (UK) grant ST/H002774/1. RM
is supported by STFC
(UK) grants ST/H002774/1 and ST/F002335/1, by a Royal Society
(UK)--NRF (South Africa) exchange grant, and by an SKA (South
Africa) Research Chair. EF thanks the ICG, Portsmouth for
hospitality during part of this work. CP thanks K. Takahashi and
K. Ichiki for kind hospitality at the University of Nagoya, and
especially for fruitful discussions on magnetic fields. RM thanks
the ACGC, University of Cape Town, where part of this work was
done, and NITHheP (South Africa) for a grant to support his visit
to ACGC.

\appendix

\section{Maxwell's equations} \label{maxeq}

Maxwell's equations $\nabla_{[\lambda}F_{\mu\nu]}=0$ and
$\nabla_\nu F^{\mu\nu}=j^\mu$ in a general spacetime take the form
\cite{Kobayashi:2007wd,Tsagas2007}
 \bea
&&{\rm D}_\mu B^\mu=- \omega_\mu E^\mu\,,~~ {\rm D}_\mu E^\mu=
\omega_\mu B^\mu+\varrho\,, \label{divbe}\\
&& \dot{B}_{\mu}^{\perp}+{2\over3}\theta B_\mu-
\big(\sigma_{\mu\nu}-\omega_{\mu\nu}\big)B^\nu \nonumber\\
&&~~~~~~= -{\rm curl}\,E_\mu-\epsilon_{\mu \nu \lambda} \dot{u}^\nu
E^\lambda \label{MaxwellBdot}\\
&& \dot{E}_{\mu}^{\perp}+{2\over3}\theta E_\mu-
\big(\sigma_{\mu\nu}-\omega_{\mu\nu}\big)E^\nu \nonumber\\
&&~~~~~~=  {\rm curl}\,B_\mu+\epsilon_{\mu \nu \lambda}
\dot{u}^\nu B^\lambda -J_\mu\,, \label{MaxwellE}
 \eea
where $E^\mu, B^\mu$ are defined by (\ref{ebfara}).
Here the total 4-current is
$j^\mu= j^\mu_{\ie}+ j^\mu_{\ip}$ and it is split as
 \be
j^\mu=\varrho u^\mu+J^\mu\,, ~~ \varrho=-u_\mu j^\mu\,, ~~
J^\mu=h^\mu_\nu j^\nu \,,
 \ee
where $\varrho,J^\mu$ are the charge density and current measured
by $u^\mu$ observers. By (\ref{svel}),
 \be \label{chnoden}
\varrho =e(\gamma_{\ip}n_{\ip}- \gamma_{\ie}n_{\ie}),~~J^\mu=
e(\gamma_{\ip}n_{\ip}v_{\ip}^\mu- \gamma_{\ie}n_{\ie}v_{\ie}^\mu).
 \ee
The derivative ${\rm D}_\mu$ is the projected covariant derivative
and it defines a covariant curl \cite{Maartens:1998xg,Tsagas2007}:
 \bea \label{deriv}
{\rm D}_\mu f &=& h^\nu_\mu \nabla_\nu f,~~ {\rm D}_\mu S^{\nu} =
h_\mu^\lambda h^\nu_\tau \nabla_\lambda S^{\tau},\\ {\rm
curl}\,S^\mu &=&\epsilon^{\mu\nu\lambda} {\rm D}_\nu S_\lambda\,.
\label{curl}
  \eea
We work in Gaussian units so that the fine structure constant is
$\alpha=e^2/(4 \pi)=1/137.036$ and the magnetic field strength is
measured in Gauss.

\section{Tetrads}

\label{tetrads}

The tetrad basis is given up to second order in scalar
perturbations by
 \bea
e_{{\zT}}{}^\mu &=& \frac{1}{a}\Big(1-\Phi+\frac{3}{2}\Phi^2\Big)
\delta^\mu_{\zT}-\frac{1}{a}S^\iB \delta_\iT^\mu \,,\\
e_{\iT}{}^\mu &=& \frac{1}{a}\Big(1+\Psi+\frac{3}{2}\Psi^2\Big)
\delta_\iT^\mu\,, \\
e^{\zT}{}_\mu &=& a\Big(1+\Phi-\frac{1}{2}\Phi^2\Big) \delta_\mu^{\zT}\,,\\
e^{\iT}{}_\mu &=& a\Big(1-\Psi-\frac{1}{2}\Psi^2\Big)
\delta^\iT_\mu+\frac{1}{a}S^\iB \delta_\mu^{\zT}\,.
 \eea
This choice of tetrad is discussed in \cite{Pitrou2008} (see also
\cite{Pitrou:2007jy,Senatore:2008vi,Durrer:1993db}). The covariant
derivative of a tensor in the tetrad basis is given by
 \be
\nabla_{\aT} X_{\bT}^{\phantom{\bT}\cT}=e_\aT^\mu\partial_\mu
X_{\bT}^{\phantom{\bT}\cT}-
\Omega_{\aT\phantom{\dT}\bT}^{\phantom{\aT}\dT}
X_{\dT}^{\phantom{\bT}\cT}+\Omega_{\aT\phantom{\cT}\dT}^{\phantom{\aT}\cT}
X_{\bT}^{\phantom{\bT}\dT}\,,
 \ee
where indices are lowered and raised with $\eta_{\aT\bT}$ and
$\eta^{\aT\bT}$. The affine connections in the background are
 \be
\bar \Omega_{\iT\zT\kT}=-\bar
\Omega_{\iT\kT\zT}=-\frac{\HH}{a}\delta_{\iT\kT}\,,~~ \HH \equiv
{a' \over a}\,,
 \ee
and the perturbed forms are
 \bea
\Omega^{(1)}_{\zT\zT\iT}&=&-\Omega^{(1)}_{\zT\iT\zT}=-\frac{1}{a}\partial_\iB
\Phi^{(1)}\,,~~ \Omega^{(1)}_{\zT\iT\kT}=0\,, \\
\Omega^{(1)}_{\iT\zT\kT}&=&-\Omega^{(1)}_{\iT\kT\zT}=\frac{1}{a}\Big[\HH
\Phi^{(1)}+\Psi^{(1)'}\Big]\delta_{\iB \kB}\,,\\
\Omega^{(1)}_{\lT\iT\kT}&=&-\Omega^{(1)}_{\lT\kT\iT}=-\frac{2}{a}\partial_{[\kB}
\Psi^{(1)}  \delta_{\iB]\ell}\,.
 \eea

\section{Euler and vorticity equations}\label{AppEuler}

\subsection{Euler equation}

For a perfect fluid with equation of state $w_\is\equiv \bar
P_\is/\bar \rho_\is$ and speed of sound $c_\is^2\equiv \dd
P_\is/\dd \rho_\is$, the term on the left of the Euler equation (\ref{geneul}) is
given to second order in the tetrad basis by
\cite{Pitrou2008,Pitrou:2008ak}:
 \bea\label{Eulerappendix} &&\frac{a \nabla_\mu
T^{\mu}_{\is\,\,\iT}}{\bar \rho_\is(1+w_\is)} =
{u^{\is}_\iT}'+(1-3 {c_\is^2})\HH
u^{\is}_\iT+\frac{c_\is^2}{1+w_\is}
\partial_\iB \delta_\is+\partial_\iB \Phi\nonumber\\
&&+\frac{1+c_\is^2}{1+w_\is}\left[(\delta_\is u^\is_\iT)'+\HH(1-3
w_\is)
\delta_\is u^\is_\iT+\delta_\is\partial_\iB \Phi\right]-4 \Psi'
u^\is_\iT\nonumber\\
&&+\partial_\jB(u^\is_\iT u_\is^\jT)-(\Phi+\Psi)\left[{u^\is_\iT}'+
\HH(1-3 c_\is^2)u^\is_\iT\right]-\partial_\iB(\Phi^2)\nonumber\\
&&+\Psi\left[{u^{\is}_\iT}'+(1-3c_\is^2)\HH
u^{\is}_\iT+\frac{c_\is^2}
{1+w_\is}\partial_\iB \delta_\is+\partial_\iB \Phi \right]\nonumber\\
&&+\frac{{c_\is^2}'}{1+w_\is}\delta_\is
u^\is_\iT-\frac{{c_\is^2}'}{3\HH(1+w_\is)^2}\,\delta_\is
\partial_\iB\delta_\is.
 \eea

\subsection{Vorticity evolution}\label{AppendixVorticity}

The vorticity tensor of species $\is$ is
 \be
\omega^\is_{\mu\nu} = h^{\is\,\alpha}_\mu h^{\is\,\beta}_\nu
\nabla_{[\alpha }u^{\is}_{\beta]}\,,
 \ee
and the vorticity vector is given by (\ref{vortdef}). In the
tetrad basis, up to second order,
 \bea
\omega^\is_\iT &=&\epsilon_{\iT \kT \lT} \omega_\is^{\kT \lT}\, \\
\label{EqExpressionomega} a \omega^\is_{\iT \kT} &=&
\partial_{[\iB} u^\is_{\kT]}+  u^\is_{[\iT} \partial_{\kB]}
(\Psi+\Phi) +u^\is_{[\iT} {u^\is_{\kT]}}'\,.
 \eea
The evolution of the vorticity is deduced from (\ref{geneul}) and
(\ref{Eulerappendix}). For a non-interacting perfect fluid, up to
second order \cite{Christopherson:2009bt,Lu:2008ju}
 \be \label{vortev}
\frac{1}{\bar \rho_\is(1+w_\is)} \partial_{[\iB}\nabla_\mu
T^{\mu}_{\is\,\,\kT]} = {\omega^\is_{\iT\kT}}'+(2-3 c_\is^2)\HH
\omega^\is_{\iT\kT}=0\,.
 \ee
This can be recast as
 \be\label{Eqadiabaticvorticity}
\big[\bar \rho_\is(1+w_\is) a^5 \omega^\is_\iT \big]'=0\,.
 \ee
For an interacting fluid,
 \bea
\hspace*{-2em}&& {\omega^\is_{\iT\kT}}'+(2-3 c_\is^2)\HH
\omega^\is_{\iT\kT}\nonumber\\ \hspace*{-2em}&&~~ ={1 \over a}
\sum_\ir\Big\{
u^\is_{[\iT}{C^{\is\ir}_{\kT]}}'+\partial_{[i}\Big(1-\Psi-{1+c_\is^2
\over 1+w_\is}\,\delta_\is \Big) C^{\is\ir}_{\kT]} \Big\}.
 \eea

\section{Magnetogenesis in tight-coupling} \label{AppBevolution}

\subsection{Magnetic field in fundamental frame}
\label{magtcff}

In the case where there are only interactions between baryons and
photons, $C_{\ib \gamma}^\mu+C_{\gamma \ib}^\mu=0$, and
 \be \label{bpcurle}
\partial_{[\iB} \nabla_\mu T^\mu_{\ib\,\kT]}+
\partial_{[\iB} \nabla_\mu T^\mu_{\gamma\,\kT]}=0\,.
 \ee
In the tight-coupled limit where the interaction rate becomes very
high, photons and baryons behave like a single fluid, with
 \be
w_\ipl=\frac{1}{3+4R}\,,~~ c^2_{s,\ipl}=\frac{1}{3(1+R)}\,, ~~
R\equiv \frac{3 \bar \rho_\ib}{4 \bar \rho_\gamma}\,.
 \ee
The energy density contrasts at first order are
 \be\label{dinTC}
\delta^{(1)}_\ipl\simeq (1+w_\ipl)\delta^{(1)}_\ib\,,\qquad
\delta^{(1)}_\ib \simeq \frac{3}{4}\delta^{(1)}_\gamma\,.
 \ee
The velocities of baryons and photons are the same in this regime
 \be\label{vinTC}
u^\ib_\iT\simeq u^\gamma_\iT \simeq u^\ipl_\iT~\Rightarrow~
\omega_\ipl^\iT\simeq \omega_\gamma^\iT\simeq\omega_\ib^\iT\,.
 \ee
By (\ref{vortev}) and (\ref{bpcurle}),
 \be\label{EvolVorticityplasma}
0\simeq {\omega^\ipl_{\iT\kT}}'+\HH(2-3 c_\ipl^2)
{\omega^\ipl_{\iT\kT}}=\frac{[\bar \rho_\ipl(1+w_\ipl) a^5
\omega^\ipl_{\iT\kT}]'}{\bar \rho_\ipl(1+w_\ipl) a^5}\,.
 \ee

This can be used to infer the source term for magnetogenesis in
(\ref{SimpleunderstandingB}). In the tight-coupled regime,
$\partial_{[\iB}\nabla_\mu T^{\mu}_{\ib\,\kT]}$ can be estimated
by using (\ref{vinTC}) and (\ref{dinTC}) in the baryon version of
(\ref{Eulerappendix}). Then, subtracting
$\partial_{[\iB}\nabla_\mu T^{\mu}_{\ipl\,\kT]}=0$, we obtain
 \bea
\hspace*{-2em}&& \frac{1}{\bar \rho_\ib}
\partial_{[\iB}\nabla_\mu T^{\mu}_{\ib\,\kT]}=3 c_\ipl^2\HH
\omega_{\iT\kT}^\ipl \nonumber\\
\hspace*{-2em}&&~
+\frac{c_\ipl^2}{a}\Big\{\frac{3\HH}{1+w_\ipl}(1-c_\ipl^2+R
c_\ipl^2)
\partial_{[\iB}\delta_\ipl v^\ipl_{\kT]}+3 \HH\partial_{[\iB}(\Psi-\Phi)
v^\ipl_{\kT]}\nonumber\\
\hspace*{-2em}&&~~~~ +\partial_{[\iB}\left(-3\Psi'+\partial_\jB
v^\jT_\ipl\right)
v^\ipl_{\kT]}-\frac{1}{1+w_\ipl}\partial_{[\iB}\Psi
\partial_{\kB]}\delta_\ipl\Big\}. \label{GenerationTermTC}
 \eea

From (\ref{SimpleunderstandingB}) it then follows that in the
tight-coupled regime, the evolution of the magnetic field is given
by
 \bea
\hspace*{-2em}&&\frac{e x_\ie}{m_\ip}
\frac{\left(a^2 B^\iT\right)'}
{a^2}=\frac{3}{2} c_\ipl^2\HH \omega^{\iT}_\ipl \nonumber\\
\hspace*{-2em}&&\qquad-\frac{c_\ipl^2}{a}\epsilon^{\iT
\lT\kT}\left\{\frac{3\HH} {1+w_\ipl}(1-c_\ipl^2+R
c_\ipl^2)\partial_{[\ell}\delta_\ipl v^\ipl_{\kT]}
\right.\nonumber\\
\hspace*{-2em}&&\qquad
\left.+\partial_{[\ell}\left(-3\Psi'+\partial_\jB
v^\lT_\ipl\right)
v^\ipl_{\lT]}-\frac{1}{1+w_\ipl}\partial_{[\lB}\Phi
\partial_{\kB]}\delta_\ipl\right\},\label{EvolBTC}
 \eea
where we used ${\rho}_\ib=(m_\ip+m_\ie)(n_\ie+ n_{\rm H})
\simeq m_\ip(n_\ie+ n_{\rm H})$.
Note that $3 c_\ipl^2\HH =\dd \ln [ \bar \rho_{\ib}/(\bar
\rho_{\ib}+4/3\bar \rho_{\gamma})]/\dd\eta$. Since the vorticity
in the tight-coupled plasma obeys (\ref{EvolVorticityplasma}), the
first term on the right hand side of (\ref{EvolBTC}), which is
linear in the vorticity, can only source the magnetic field if
there is initially vorticity in the plasma. This is the term
responsible for the Harrison mechanism
\cite{1970MNRAS.147..279H,Hollenstein:2007kg}. All other terms
which are quadratic can source the magnetic field even if there is
no initial vorticity.

However, {\em on large scales in the radiation era there is a
suppression of the total contribution of these quadratic terms}.
From the large-scale radiation era relations at first order,
 \be
2\HH \partial_\iB v_\ipl^\iT\simeq\nabla^2
\Phi\,,\qquad\delta_\ipl\simeq-2 \Phi\,,
 \ee
it follows that at lowest order the quadratic terms are estimated
by $\partial_i X
\partial_j Y  \sim \partial_i \Phi \partial_j\Phi $.
Hence the quadratic source terms are suppressed by a factor
$(k\eta)^2$, since at lowest order all contributions are of the
type $\sim
\partial_{[\iB}\Phi \partial_{\jB]}\Phi=0$. This implies that
$\sqrt{k^3 P_B(k,\eta)}\propto k^4 \eta^2/\eta_\eq^2$, that is
$\sum_i {\cal S}_i\propto k^3 \eta^5/\eta_\eq^2$.

\subsection{Magnetic field in baryon frame}
\label{AppBevolutionBaryonsFrame}

From (\ref{BChgFrame}) we obtain
 \bea
B^\iT_\ib-B^\iT&=& -\epsilon^{\iT \lT \kT}v^\ib_\lT
E_\kT=-\frac{1}{ e (  n_\ie+  n_{\rm H}) x_\ie}\epsilon^{\iT \lT \kT}v^\ib_\lT
\nabla_\mu T_{\ib\,\kT}^\mu \nonumber\\
&=&\frac{m_\ip}{a e
x_\ie}\frac{c_\ipl^2}{(1+w_\ipl)}\epsilon^{\iT \lT \kT}v^\ipl_\lT
\partial_{\kT} \delta_\ipl\,,
 \eea
where the second equality holds in the tight-coupled regime. Using
the first order version of the Euler equation
(\ref{Eulerappendix}) for the plasma, i.e. with $\nabla_\mu
T^{\mu\iT}_\ipl=0$, and using also the first order evolution
equation for the plasma density contrast,
 \be
\Big(\frac{\delta_\ipl}{1+w_\ipl} \Big)' = 3\Psi' -
\partial_\iB v^\iT\,,
 \ee
we deduce that in the tight-coupled regime
 \be
\frac{e  x_\ie}{m_\ip} \frac{\big(a^2
B^\iT_\ib\big)'}{a^2}=3 c_\ipl^2\HH
\omega^{\iT}_\ipl=-\frac{\big(a^2 \omega^{\iT}_\ipl
\big)'}{a^2}\,.
 \ee
%
At early times in the radiation era we have $x_\ie\simeq 1$, and then we
obtain a conservation equation up to second order:
 \be
\Big[a^2 \Big(
\frac{e}{m_\ip}B^\iT_\ib+{\omega^{\iT}_\ipl}\Big)\Big]'\simeq 0\,.
 \ee
This is precisely the Harrison mechanism, but up to second order.

In the tight-coupled regime, in the plasma frame, the magnetic
field can only be generated if there is initial vorticity, i.e.
through the Harrison mechanism. We recover here the results in
\cite{Kobayashi:2007wd,Maeda:2008dv}. The magnetic field measured
in a different frame is only due to the contribution of the
electric field to this change of frame.  In the fundamental frame,
this contribution in the tight-coupled regime is given by the
second and third lines of (\ref{EvolBTC}). Note that the electric
field is generated at first order in cosmological perturbations
even in the lowest order of the tight-coupling approximation and
even in the plasma frame.



\end{document}